\documentclass[10pt,journal,compsoc]{IEEEtran}

\IEEEoverridecommandlockouts

\usepackage[T1]{fontenc}
\usepackage{cite}
\usepackage{hyperref}
\usepackage{amsmath,amssymb,amsfonts,amsthm}
\usepackage{algorithmic}
\usepackage{graphicx}
\usepackage{textcomp}
\usepackage{xcolor}
\usepackage{braket}
\usepackage{lipsum}
\usepackage{tikz}
\usepackage{quantikz}
\usepackage{float}
\usepackage{subcaption}
\usepackage{array}
\usepackage{algorithm2e}
\usepackage{xfrac}
\usepackage{mdframed}
\RestyleAlgo{ruled}
\SetKwComment{Comment}{/* }{ */}

\usetikzlibrary{positioning}

\def\BibTeX{{\rm B\kern-.05em{\sc i\kern-.025em b}\kern-.08em
    T\kern-.1667em\lower.7ex\hbox{E}\kern-.125emX}}

\newtheorem{definition}{Definition}
\newtheorem{lemma}{Lemma}

\newtheorem*{graphstate}{Graph State}
\newtheorem*{lcequivalence}{LC equivalence}
\newtheorem*{paulimeasurements}{Projective Measurements via Pauli Operators}
\newtheorem*{goal}{Research Problem}
\newtheorem*{initialstate}{Elementary State}
\newtheorem*{designprinciples}{Design Principles}
\newtheorem*{chaingraph}{Chain Graph State}
\newtheorem*{treelikegraph}{Generalized Tree-Like Graph State}

\theoremstyle{remark}
\newtheorem*{remark}{Remark}

\newcommand{\eqdef}{\stackrel{\triangle}{=}}

\begin{document}

\title{Intra-QLAN Connectivity: \\beyond the Physical Topology}

\author{\IEEEauthorblockN{Francesco Mazza, Marcello~Caleffi,~\IEEEmembership{Senior~Member,~IEEE}, Angela~Sara~Cacciapuoti,~\IEEEmembership{Senior~Member,~IEEE}}
    \thanks{The conference version \cite{MazCalCac-24} of this paper has been accepted in the Proc. of IEEE QCNC'24.}
    \thanks{The authors are with the \href{www.quantuminternet.it}{www.QuantumInternet.it} research group, \textit{FLY: Future Communications Laboratory}, University of Naples Federico II, Naples, 80125 Italy. A.S. Cacciapuoti and M. Caleffi are also with the Laboratorio Nazionale di Comunicazioni Multimediali, National Inter-University Consortium for Telecommunications (CNIT), Naples, 80126, Italy. }
    \thanks{Angela Sara Cacciapuoti and Francesco Mazza acknowledge PNRR MUR NQSTI-PE00000023, Marcello Caleffi acknowledges PNRR MUR project RESTART-PE00000001.}
}
\maketitle

\begin{abstract}
    In the near to mid future, Quantum Local Area Networks (QLANs) -- the fundamental building block of the Quantum Internet -- will unlike exhibit physical topologies characterized by densely physical connections among the nodes. On the contrary, it is pragmatic to consider QLANs based on simpler, scarcely-connected physical topologies, such as star topologies. This constraint -- if not properly tackled -- will significantly impact the QLAN performance in terms of communication delay and/or overhead. Thankfully, it is possible to create on-demand links between QLAN nodes, without physically deploying them, by properly manipulating a shared multipartite entangled state. Thus, it is possible to build an overlay topology, referred to as \textit{artificial topology},  upon the physical one. In this paper, we address the fundamental issue of engineering the artificial topology of a QLAN to bypass the limitations induced by the physical topology. The designed framework relays only on local operations, without exchanging signaling among the QLAN nodes, which, in turn, would introduce further delays in a scenario very sensitive to the decoherence. Finally, by exploiting the artificial topology, it is proved that the troubleshooting is simplified, by overcoming the single point of failure, typical of classical LAN star topologies.
\end{abstract}
    

\section{Introduction}
\label{sec:1}

Interconnecting different quantum processors with a Quantum Local Area Network (QLAN) -- namely, with a quantum network able to cover a limited geographic area -- represents one of the very first steps for unlocking the vision of the Quantum Internet \cite{CacCalTaf-20,DurLamHeu-17,IllCalMan-22,PirDur-18,RamPirDur-21}. And trial deployments of quantum server farms based on technologies mimicking QLAN have already begun \cite{IBM2025,AWSQN,KnaSulWei-24, LiuJiaLuo-24, StoKiaMar-24}.

It must be noted, though, that current (and near-term) state-of-the-art of QLAN hardware requires sophisticated and resource-intensive setups, often involving complex experimental apparatuses and precise control mechanisms. Thus, in the short-mid time horizon, physical topologies like fat tree and leaf-spine, typical of classical data centers \cite{LebManTiw-16, YaoWuVen-14} and characterized by densely physical connections among the LAN nodes, are not practical in QLANs. Conversely, it is quite reasonable and pragmatic to consider simpler physical topologies for QLANs, such as star topologies \cite{ChuRamAni-24}, characterized by weaker connectivity among the QLAN nodes.

Yet, these constraints on the physical topology induced by the quantum hardware underlying QLAN functioning -- if not properly tackled --  impact and limit the achievable communication capabilities among the QLAN nodes. 

To this aim, we cannot borrow well-established approaches from classical LANs. In fact, regardless of the particulars of the physical (classical) LAN topology, upper layers of the classical protocol stack are responsible to overcome the constraints imposed by the physical topology enabling any-to-any communication, at the price of communication overhead and information duplication. Clearly, these are not viable strategies in QLAN due to unconventional quantum peculiarities, ranging from  stringent coherence times to quantum mechanics postulates and phenomena, such as quantum measurement and the no-cloning theorem. Besides, the design of a protocol suite for quantum networks is still at its infancy \cite{IllCalMan-22}, and thus the functionalities of ``quantum'' upper-layers are yet to be defined.  

Fortunately, the communication limitations induced by the physical topology in QLANs can be overcame by relying on the most distinguish feature of quantum mechanics, namely, quantum entanglement. Specifically, entanglement enables a new and richer form of connectivity \cite{IllCalMan-22,CacIllCal-23,CheIllCac-24}, referred in the following as \textit{entanglement-enabled connectivity}, with no-counterpart in classical LANs.

In fact, once an entangled state -- say an EPR pair for the sake of exemplification -- has been shared between two nodes, a qubit can be ``transmitted'' via quantum teleportation \cite{BenBraCre-93,CacCalVan-20}, which does not require the physical transmission of the quantum particle encoding the qubit on the physical channel. Accordingly, entanglement enables \textbf{half-duplex unicast channels} between any pairs of nodes, regardless of their relative positions within the \textbf{underlying physical network topology}. Hence, QLAN nodes that are not physical connected can still directly communicate, as long as they share some entanglement.

Additionally, entanglement is not limited to EPR pairs. With multipartite entanglement \cite{DurVidCir-00,EisBriHan-01}, the nature of the entanglement-enabled connectivity becomes richer. As instance, by distributing an $n$-qubit GHZ state \cite{DurVidCir-00} among $n$ QLAN nodes, an EPR pair can be distributively extracted by any pair of nodes, with the identities of the entangled nodes chosen at run-time, accordingly to the traffic demand. Furthermore, changing the selected multipartite entangled state changes the specific communication patterns enabled by the \textit{entanglement-enabled connectivity}. 

Consequently, by exploiting multipartite entanglement, it is possible \textbf{to augment the physical topology by introducing \textit{artificial links} between un-connected nodes, without any additional physical link}. It may be useful to clarify that an artificial link between two QLAN nodes reflects the interaction pattern between the qubits belonging to the composite multipartite entangled state. Thus, an artificial link denotes the ``possibility'' of extracting a shared EPR between the two nodes, starting from a multipartite entangled state shared among a larger set of nodes. However, the number of EPR pairs that can be simultaneously extracted from a single multipartite entangled state heavily depends on the type and structure of the considered state \cite{HeiDurEis-06}, and some of the artificial links are depleted during the extraction process.

These artificial links constitute a sort of ``overlay entangled topology'' built upon the physical one,  referred to as \textit{artificial topology}, that can differs significantly from the physical topology.

In this paper, we shed the lights on the possibility offered by engineering the artificial QLAN topology to overcome the limitations and the communication constraints induced by the physical QLAN topology. This possibility has no counterpart in classical networks. Indeed, it is interesting to observe that, since entanglement is widely recognized as a communication resource reminiscent of a resource encompassing both the classical physical and link layers \cite{IllCacMan-21,KozWehVan-22}, our findings highlight that the capability to enable any-to-any communication is not constrained to be delegated to the upper layers of the eventually designed protocol stack for the Quantum Internet. Furthermore, through the paper, we show how  is possible to engineer the artificial topology accordingly to the on-demand traffic requests. And our framework relies only on local operations, without exchanging signaling among the QLAN nodes, which, in turn, would introduce further delays in a scenario very sensitive to the decoherence.

We finally prove that, by exploiting the artificial topology, it is possible to simplify the troubleshooting, by overcoming the single point of failure, typical of classical LAN star topologies.
 
\subsection{Related Work}
\label{sec:1.1}

In this manuscript, we exploit the properties of a class of multipartite entangled states, referred to as \textit{graph states} \cite{HeiEisBri-04}, that recently gained significant attention from the community \cite{WalZweMus-16, Tzitrin-18, BarBirBom-23, BenHajVan-23, LeeJeo-23} due to their unique entanglement properties \cite{HeiDurEis-06, HeiEisBri-04}. Indeed such properties make graph states ideal resources for various applications in quantum computing and quantum communications \cite{CacIllCal-23,RamPirDur-21}.

In particular, graph states \cite{HeiDurEis-06} have been widely investigated within the context of fusion-based strategies for building larger states, starting from smaller building blocks \cite{ShaSha-23,LeeJeo-23}. Graph states have also been extensively studied for ``routing'' EPRs through network nodes \cite{HahPapEis-19}. And, the measurement-based properties \cite{HeiDurEis-06} of graph states have been explored for repeating operations within the optical implementation of quantum repeaters \cite{BenHajVan-23}.

Differently from the mentioned literature, in this paper the aim is to dynamically adapt the artificial topology, associated to an initial graph state, to the QLAN traffic patterns, by overcoming the limitations induced by the physical topology. 

To the best of our knowledge, this is the first paper engineering the intra-QLAN connectivity, by exploiting entanglement. And, indeed, there was an urgent call for doing this. In fact, recently in \cite{CheIllCac-24}, it has been shown that the inter-QLANs connectivity, namely, the connectivity among different QLANs, heavily depends on the multipartite entangled states locally generated and distributed within the single QLANs. 

\vspace{6pt}
The remaining part of the manuscript is organized as follows. In Sec.~\ref{sec:2}, we introduce some preliminaries related to graph states. In Sec.~\ref{sec:3}, we first describe the system model, and then in Sec.~\ref{sec:4}, we develop the theoretical analysis by providing the tools for engineering the artificial topology beyond the limitations induced by the physical one. Finally, in Sec.~\ref{sec:5}, we conclude the paper.

\section{Preliminaries}
\label{sec:2}

In this section, we first overview some preliminaries related to graph theory in Sec.~\ref{sec:2.1}, which are used in Sec.~\ref{sec:2.2} to present and describe the class of multipartite entangled states exploited in the remaining part of the manuscript -- namely, \textit{graph states} -- and the main tools for their manipulation.

\subsection{Graph theory fundamentals}
\label{sec:2.1}

Formally, a graph $G$ is defined as a pair of two (finite) sets, $V$ and $E$, as follows:
\begin{equation}
	\label{eq:01}
	G \eqdef (V,E),
\end{equation}
with $V$ denoting the set of vertices -- also called \textit{nodes} -- with cardinality $|V| = n$, and $E$ denoting the set of edges describing the connections between the vertices:
\begin{equation}
	\label{eq:02}
	E = \left\{ \{a,b\}: a,b \in V \wedge a \neq b \right\}\subset V \times V \eqdef V^2
\end{equation} 
In the following we utilize the compact notation introduced in \eqref{eq:02} also for two arbitrary vertex sets $A,B \subseteq V$, by denoting with the symbol $A \times B \subseteq V^{2}$ the set of all possible edges having one endpoint in $A$ and the other in $B$:
\begin{equation} 
	\label{eq:03}
	A \times B \eqdef \{ \{a,b\}\in V^{2}: a\in A \wedge b \in B \}.
\end{equation}

\begin{remark}
	Accordingly to the above, we have restricted our attention on finite graphs, i.e., graphs with finite set of vertices and edges. Furthermore, from \eqref{eq:02}, we have considered only \textit{undirected} and \textit{simple} graphs (i.e., graphs where an edge cannot connect the same vertex) since these two properties are required for the mapping between graphs and graph states \cite{HeiEisBri-04,HeiDurEis-06}. Indeed, the three mentioned properties are quite reasonable from a merely network perspective, where real-world networks always interconnect a finite set of nodes and a link having the same node as start and end-point has no usefulness from a communication perspective.
\end{remark}

Among the graph properties, vertex adjacency will be heavily used through the paper, as evident from the next definitions. Formally, given two vertices $a,b \in V$, if $a$ and $b$ are connected through an edge $\{a,b\}\in E$, then they are defined as \textit{adjacent} vertices. 

\begin{definition}[\textbf{Open and Closed neighborhood}]
	\label{def:01}
	The set $N_a$ of vertices adjacent to an arbitrary vertex $a$ is called \textit{open neighborhood} of $a$, and it is defined as: 
	\begin{equation}
		\label{eq:04}
		N_a \eqdef \{b \in V : \{a,b\} \in E\}.  
	\end{equation}
	The term ``open'' highlights that the vertex $a$ is not included in the set. Accordingly, a vertex $a$ such that $|N_a| = 0$ is called \textit{isolated vertex}. Conversely, whenever the vertex $a$ should be included as well within the set, we utilize the ``closed" neighborhood $\dot{N}_a$ of $a$:
	\begin{equation}
		\label{eq:05}
		\dot{N}_a \eqdef N_a \cup \{a\}.  
	\end{equation}
\end{definition}

\begin{definition}[\textbf{Induced Subgraph}]
	\label{def:02}
	The subgraph of $G = (V,E)$ induced by a vertex set $A \subseteq V$ is defined as the graph $G[A]$ having: i) as vertices, the ones in $A$, and ii) as edges, the edges in $E$ whose endpoints are both in $A$. Formally:
   \begin{equation}
	   \label{eq:06}
	   G[A] \eqdef (A, E_{A}),
   \end{equation}
   with:
   \begin{equation}
	   \label{eq:07}
	   E_{A}\eqdef \big\{\{b,c\} \in E: b \in A \wedge c \in A \big\} = E \cap A^2.
   \end{equation}
	Whether $A$ should coincide with the neighborhood $N_a$ of some vertex $a$, then $G[N_a]$ is referred to as the \textit{subgraph induced by the neighborhood of $a$}.
\end{definition}

\begin{definition}[\textbf{Complete graph}]
	\label{def:03}
	A complete graph, also referred to as fully connected graph, is a graph where each pair of the $n = |V|$ vertices is adjacent. Formally:
	\begin{equation} 
		\label{eq:08}
		K_n \eqdef (V, V^{2}) 
	\end{equation}
\end{definition}

\begin{definition}[\textbf{Star vertex}]
	\label{def:04}
	A star vertex of graph $G=(V,E)$, also referred to as completely connected vertex, is a vertex $s \in V$ adjacent to all the other vertices in $V \setminus \{ s \}$. The set $S$ of star vertices of G is given by: 
	\begin{equation} 
		\label{eq:09}
		S \eqdef \{s \in V: N_{s} = V \setminus \{s\} \} \subseteq V.
	\end{equation}
\end{definition}

Clearly, from \eqref{eq:09} it follows that, given a star vertex $s$ of a graph $G=(V,E)$, its closed neighborhood set $\dot{N}_s$ coincides with $V$. However, in the following we use the notation $\dot{N}_s$ whenever we want to emphasise the role played by $s$. 

\begin{definition}[\textbf{Induced star subgraph}]
	\label{def:05}
	Let $s \in V$ be a star vertex of a graph $G=(V,E)$. The subgraph of $G$ induced by its closed neighborhood $\dot{N}_s$ is referred to as induced star subgraph $G[\dot{N}_s]$:
	\begin{equation}
		\label{eq:10}
		G[\dot{N}_s] \eqdef (\dot{N}_s, \{s\} \times N_s).
	\end{equation}
	Accordingly, $s$ is the star vertex of the subgraph $G[\dot{N}_s]$.
\end{definition}

\begin{definition}[\textbf{Graph complementation}]
	\label{def:06}
	The complement (or inverse) of a graph $G$ is the graph $\tau(G)$, obtained by considering the same set of vertices $V$ but with the edge set built such that two distinct vertices of $\tau(G)$ are adjacent if and only if they are not adjacent in $G$. Formally:
	\begin{equation}
		\label{eq:11}
		\tau(G) = (V, E^C),
	\end{equation}
	with 
	\begin{equation}
		\label{eq:12}
		E^C\eqdef V^2 \setminus E = \{ \{a,b\} \in V^2 : \{a,b\} \not\in E\}.	
	\end{equation}
\end{definition}

\begin{figure*}[t]
	\centering
	\begin{subfigure}{0.45\textwidth}
		\centering
		\begin{tikzpicture}[scale=0.85]
  % Define a common style for the 3D appearance
  \tikzstyle{node3d} = [circle, draw=none, shading=ball, text=white, ball color=blue!70]

  % Nodes with 3D appearance
  \node[node3d] (A) at (0,0) {a};
  \node[node3d] (B) at (1.5,0) {b};
  \node[node3d] (C) at (3,0) {c};
  \node[node3d] (D) at (4.5,0) {d};
  \node[node3d] (E) at (6,0) {e};

  % Edges
  \draw (A) -- (B);
  \draw (B) -- (C);
  \draw (C) -- (D);
  \draw (D) -- (E);
\end{tikzpicture}
		\caption{Representation of the graph $G$ associated to a 5-qubits \textit{linear} graph state $\ket{G}$.}
	\end{subfigure}
	\hfill
	\vspace{0.2cm}
	\begin{subfigure}{0.5\textwidth}
		\centering
		\begin{tikzpicture}[scale=0.85]

  \tikzstyle{node3d} = [circle, draw = none, shading=ball, text=white, ball color=blue!70]
  
  \node[node3d] (A) at (0,0) {a};
  \node[node3d] (B) at (1.5,0) {b};
  \node[node3d] (C) at (3,0) {c};
  \node[node3d] (D) at (4.5,0) {d};
  \node[node3d] (E) at (6,0) {e};
  \draw (A) -- (B);
  %\draw (B) -- (C);
  %\draw (C) -- (D);
  \draw (D) -- (E);
  %\draw (C) to[bend left = 25] (E);
\end{tikzpicture}
		\caption{Representation of the graph ${\Tilde{G}_z = G - c}$ obtained by performing a $\sigma_z$-measurement on the qubit associated to vertex $c$.}
	\end{subfigure}
	\vspace{0.2cm}
	\begin{subfigure}{0.45\textwidth}
		\centering
		\begin{tikzpicture}
    \tikzstyle{node3d} = [circle, draw = none, shading=ball, text=white, ball color=blue!70]

    \node[node3d] (A) at (0,0) {a};
    \node[node3d] (B) at (1.5,0) {b};
    \node[node3d] (C) at (3,0) {c};
    \node[node3d] (D) at (4.5,0) {d};
    \node[node3d] (E) at (6,0) {e};
    \draw (A) -- (B);
    \draw (D) -- (E);
    \draw (D) to[bend right = 25] (B);
  \end{tikzpicture}
		\caption{Representation of the graph ${\Tilde{G}_y = \tau_c(G)-c}$ obtained by performing a $\sigma_y$-measurement on the qubit associated to vertex $c$.}
	\end{subfigure}%
	\hfill
	\vspace{0.2cm}
	\begin{subfigure}{0.52\textwidth}
		\centering
		\begin{tikzpicture}[scale=0.85]
    \tikzstyle{node3d} = [circle, draw = none, shading=ball, text=white, ball color=blue!70]

    \node[node3d] (A) at (0,0) {a};
    \node[node3d] (B) at (1.5,0) {b};
    \node[node3d] (C) at (3,0) {c};
    \node[node3d] (D) at (4.5,0) {d};
    \node[node3d] (E) at (6,0) {e};
    \draw (A) -- (B);
    %\draw (D) -- (E);
    \draw (D) to[bend right = 25] (B);
    \draw (B) to[bend right = 25] (E);
  \end{tikzpicture}
		\caption{Representation of the graph ${\Tilde{G}_x = \tau_d(\tau_c( \tau_d(G)-c))}$ obtained by performing a $\sigma_x$-measurement on the qubit associated to vertex $c$.}
	\end{subfigure}
	\caption{Pictorial representation of the effects of different single-qubit Pauli-measurements on a graph state. The effects are shown by representing the graph associated with the graph state obtained after the measurements (up to local unitaries). As widely done, a graph is represented by a diagram in a plane, where vertexes are denoted by points in the plane and edges are denoted by arches between two vertices. }
	\hrulefill
	\label{fig:01}
\end{figure*}
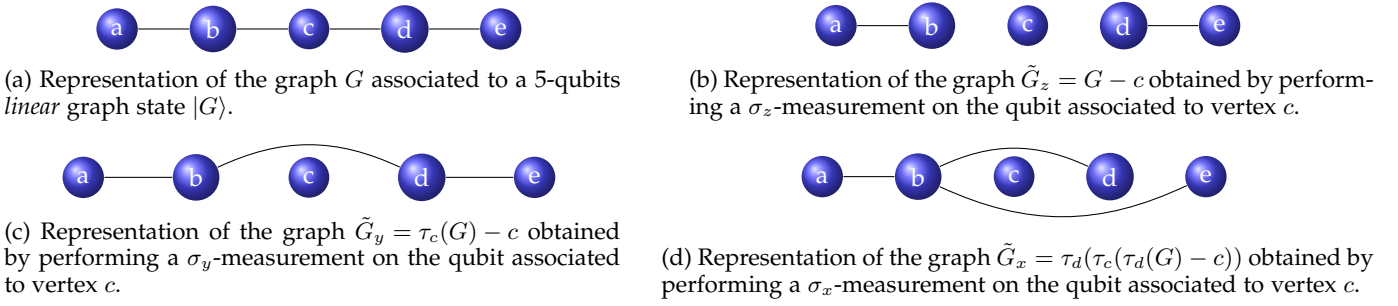

The complementation can be done also with respect to the subgraph $G[N_a]$ induced by the neighborhood $N_a$ of a vertex $a \in V$. In this case, it is usually referred as \textit{local complementation} of $G$ at vertex $a$, and it is denoted as $\tau_a(G)$, as formally defined below. 

\begin{definition}[\textbf{Local Complementation}]
	\label{def:07}
	Given a graph $G=(V,E)$, the local complementation of $G$ at vertex $a \in V$ is the graph $\tau_a(G)$ obtained by complementing the subgraph $G[N_a]$ induced by neighborhood $N_a$ of vertex $a$, while leaving the rest of the graph unchanged:
	\begin{align}
		\label{eq:13}
		\tau_a(G) = \big( V, ( E \cup N_a^2 ) \setminus E_{N_a}\big)
	\end{align}
	with $E_{N_a}$ defined in Def.~\ref{def:02}.
\end{definition}

\begin{definition}[\textbf{Vertex deletion}]
	\label{def:08}
	Given a graph $G=(V,E)$, the deletion of a vertex $a \in V$ generates a new graph, denoted as $G - a$, where both vertex $a$ and all the edges connecting $a$ with its adjacent vertexes are removed. Formally:
\begin{equation}
	\label{eq:14}
	G - a = \big(V \setminus \{a\}, E \setminus ( \{a\} \times N_a ) \big)
\end{equation}
Hence, the edge set of $G - a$ is the set of edges in $G$ without the edges with vertex $a$ as endpoint.
\end{definition}

\begin{definition}[\textbf{Path}] 
	\label{def:09}
	A $\{a,b\}$-path is an ordered list $p_{\{a,b\}} \eqdef (a_1,a_2,\ldots,a_l)$ of distinct vertices in $V$ so that $a=a_1$, $b = a_l$ and  $\{a_i, a_{i+1}\} \in E$ for any $i$. 
\end{definition}
Accordingly, a graph $G=(V,E)$ is \textit{connected} if, for each pair of vertices $a,b \in V$, there exists a $\{a,b\}$-path in E.

\subsection{Multipartite Entanglement: Graph states} 
\label{sec:2.2}

A notable class of multipartite entangled states from a communication perspective is represented by the so-called \textit{graph states} \cite{HeiEisBri-04,HeiDurEis-06}, which -- as suggested by the name -- can be effectively described with the graph theory tools introduced in Sec.~\ref{sec:2.1}. Specifically, stemming from an arbitrary graph $G$ defined in \eqref{eq:01}, the corresponding \textit{graph state} $\ket{G}$ is obtained by mapping each vertex of the graph $G$ with a qubit in the state $\ket{+}$, and then performing a controlled-Z (\texttt{CZ}) gate between each pair of qubits corresponding to adjacent vertices in $G$. The rationale underlying such a mapping lies in the correspondence between graph edges and interaction patterns among the qubits belonging to the composite entangled system. In the mapping, vertices play the role of physical systems and edges represent their interactions.

\begin{graphstate}
	Formally, the $n$-qubit \textit{graph state} $\ket{G}$ associated to graph $G \eqdef (V,E)$ can be expressed\footnote{With a (widely adopted) notation abuse, since the application of the $\texttt{CZ}_{ab}$ gate on the state $\ket{+}^{\otimes n}$ requires a reference to $n-2$ identity operations $I$ acting on all the qubits different from $a$ or $b$.} as \cite{HeiDurEis-06}:
	\begin{equation}
		\label{eq:15}
		\ket{G} = \prod_{\{a,b\} \in E} \texttt{CZ}_{ab}\ket{+}^{\otimes n},
	\end{equation}
	with $\ket{+} = \tfrac{1}{\sqrt{2}}(\ket{0} + \ket{1})$, $n = |V|$ and $\texttt{CZ}_{ab}$ denoting the $\texttt{CZ}$ gate applied to the qubits associated to the vertices $a$ and $b$.
\end{graphstate}

One could wonder whether graph states associated to different graphs may be equivalent up to some metric. This is clarified by the following definition.

\begin{definition}[\textbf{LU equivalence}]
	\label{def:10}	
	Given two $n$-qubit quantum states, say $\ket{G}$ and $\ket{G'}$, then $\ket{G}$ and $\ket{G'}$ are Local-Unitary (LU)-equivalent \textit{iff} there exists $n$ local-unitary operators $\{ U_i \}$ so that \cite{Kra-10}:
	\begin{equation}
		\label{eq:16}
		\ket{G}=\bigotimes_i U_i \ket{G}
	\end{equation}
\end{definition}

Accordingly to the above definition, it results that -- although each graph state $\ket{G}$ corresponds uniquely to a graph $G$ -- graph states associated to different graphs might be equal up to some LU-operations \cite{HeiEisBri-04, HeiDurEis-06}. The mapping between graph states and graphs is crucial beyond a merely pictorial purpose. Specifically, the action of key operations on a graph state $\ket{G}$ can be described via simple transformations on the associated graph G. $A$mong the possible operations on graph states, Local-Clifford (LC) unitaries (which are a subset of Local-Unitary operators) \cite{HeiDurEis-06} and single-qubit Pauli measurements play a crucial role for the objectives of this manuscript.

Regarding LC unitaries, their actions can be described via local complementations defined on the corresponding graph. Indeed, the following result holds \cite{HeiDurEis-06}. 

\begin{lcequivalence}
	Two $n$-qubit quantum states, say $\ket{G}$ and $\ket{G'}$ are LC-equivalent iff the corresponding graphs $G$ and $G'$ are related by a sequence of local complementations defined in Def.~\ref{def:07}.
\end{lcequivalence} 

\begin{figure*}[t]
	\centering
	\begin{subfigure}{0.39\textwidth}
		\centering
		\includegraphics[width=0.85\textwidth]{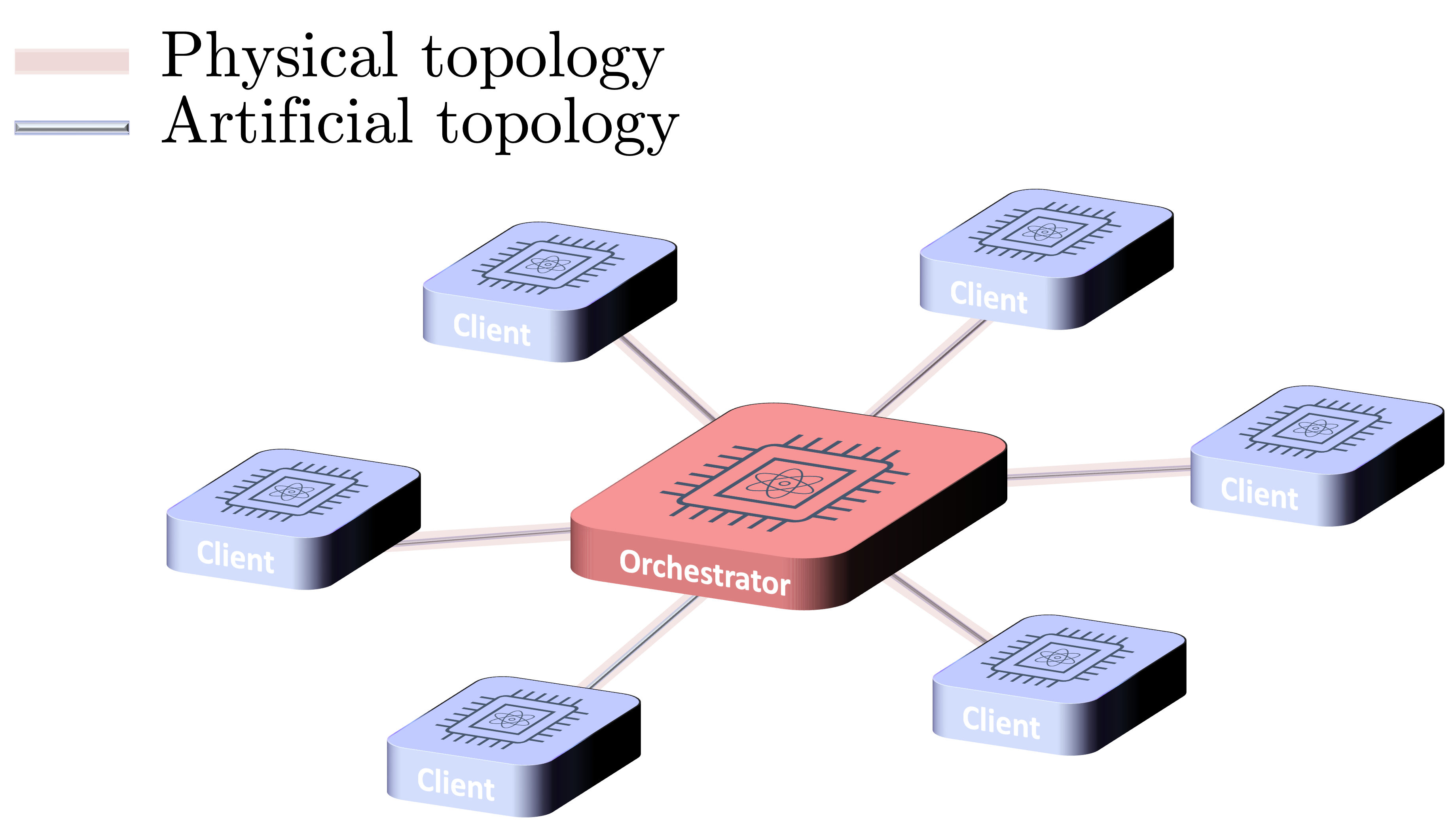}
		\caption{}
	\end{subfigure}%
	\hfill
	\begin{subfigure}{0.28\textwidth}
		\centering
		\includegraphics[width=\textwidth]{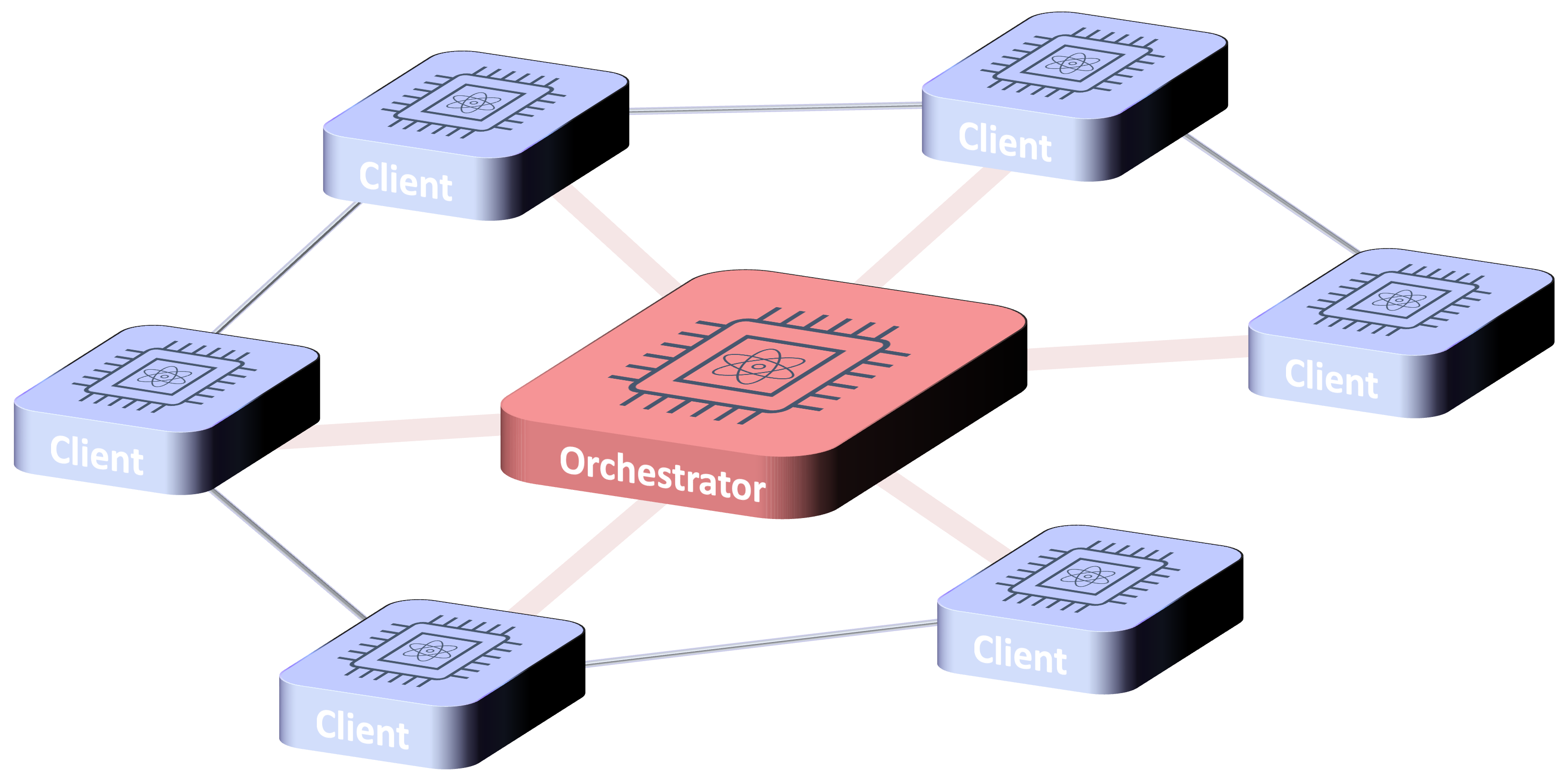}
		\caption{}
	\end{subfigure}
	\hfill
	\begin{subfigure}{0.28\textwidth}
		\centering
		\includegraphics[width=\textwidth]{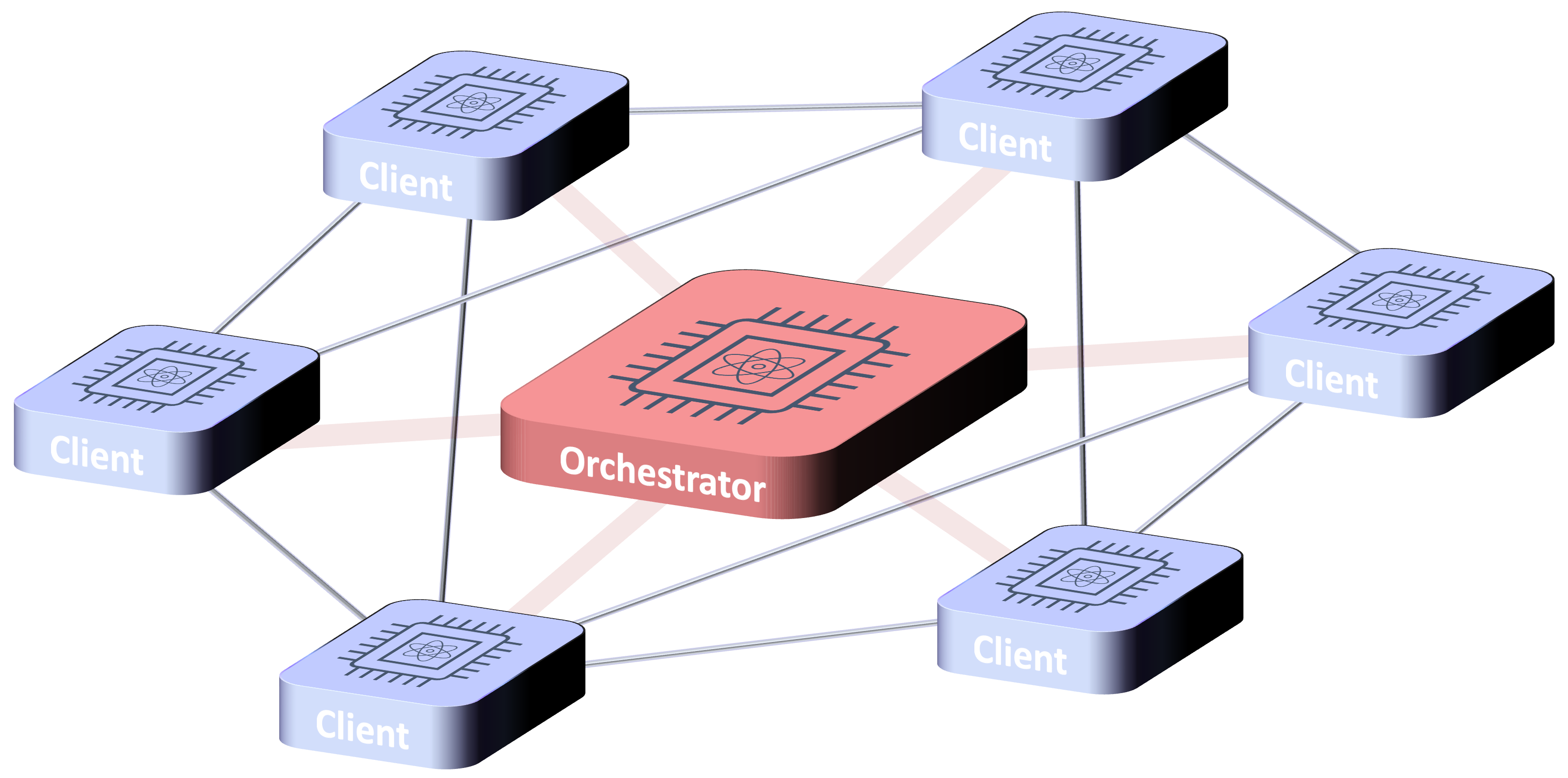}
		\caption{}
	\end{subfigure}
	\caption{Pictorial representation of a QLAN. The orchestrator node (shown in red) is connected to the client nodes via a physical topology. After operations performed locally at the orchestrator, artificial topologies are built upon the physical one: artificial bus topology, in the sub-figure (b) or artificial (enhanced) ring topology, in the sub-figure (c).}
	\hrulefill
	\label{fig:02}
\end{figure*}

Regarding single-qubit Pauli measurements, a projective measurement through a Pauli operator $\sigma_x, \sigma_y,$ or $\sigma_z$ on a qubit of the graph state $\ket{G}$ yields, up to local unitaries $ U_{i,\pm}$, a new graph state $\ket{\tilde{G}}$ on the unmeasured qubits. Interestingly, as proved in \cite{HeiDurEis-06,HeiEisBri-04}, this new graph state $\ket{\tilde{G}}$ can be obtained by means of vertex deletion and/or local complementation (Defs.~\ref{def:07} and \ref{def:08}) on the graph $G$ associated to the original graph state $\ket{G}$, as summarized in the following and represented in Fig.~\ref{fig:01}. 

\begin{paulimeasurements}
	The projective measurement of a qubit -- associated to vertex $a \in V$ in graph $G = (V,E)$ -- of the initial graph state $\ket{G}$ through a Pauli operator $\sigma_\chi$ yields, up to local unitaries, to a new graph state $\ket{\tilde{G}_\chi}$\footnote{With a mild notation abuse, the dependence on $a$ is neglected for the sake of notation simplicity. Similar notation abuses will be adopted also for the following projective measurements.} among the remaining qubits, whose associate graph $\tilde{G}_\chi$ is obtained:
	\begin{itemize}
		\item[-] for Pauli operator $\chi = \sigma_z$, by deleting the vertex $a$ from graph $G$:
			\begin{equation} 
				\label{eq:17}
				\tilde{G}_z \eqdef G - a.
			\end{equation}
		\item[-] for Pauli operator $\chi = \sigma_y$, by first local complementation of the graph $G$ at vertex $a$, and then by deleting $a$ from graph $G$:
			\begin{equation} 
				\label{eq:18}
				\tilde{G}_y\eqdef \tau_a(G) - a. 
			\end{equation}
		\item[-] for Pauli operator $\chi = \sigma_x$, by concatenating the following three graph operations: $i)$ local complementation of the graph $G$ at an arbitrary neighbour vertex $b_0 \in N_a$, $ii)$ then, local complementation of the graph $G$ at vertex $a$, followed by the deletion of $a$ from graph $G$, and $iii)$ finally, a local complementation at $b_0$ of the graph obtained at the previous step:
			\begin{equation} 
				\label{eq:19}
				\tilde{G_x} = \tau_{b_0}\big( \tau_{a} (\tau_{b_0}(G)) - a\big).
			\end{equation}
			It is worthwhile to note that, although the choice of the vertex $b_0$ in the neighborhood of $a$ at step $i)$ is not unique, the post-measurement graph states are LU equivalent for any choice of $b_0$ \cite{HeiDurEis-06}.
		\end{itemize}
\end{paulimeasurements}

\section{Model and Design Parameters} 
\label{sec:3}

In this section, we exploit the tools introduced in Sec.~\ref{sec:2} to show how the artificial QLAN topology can be engineered from a communication perspective to overcome the communication constraints induced by the physical QLAN topology. Accordingly, we have the following research objective.

\begin{goal}
Our goal is to create direct, artificial links between nodes that are not neighbors in the physical topology, so that they can directly communicate, overcoming so the communication limitations induced by the physical topology. Furthermore, we aim at creating such links on-demand -- i.e., at run-time, whenever needed -- so that these direct links can adapt to varying communication needs, by properly engineering a multipartite entanglement state shared via the physical topology.
\end{goal}

\subsection{System Model}
\label{sec:3.1}

Entanglement generation and distribution within a QLAN require highly specialized environments equipped with challenging and expensive hardware -- as instance ultra-high vacuum systems or ultra-low temperature cryostats -- necessary to preserve the coherence of the quantum states. And, the challenges for controlling and preserving the quantum states get harder as the number of physical connections increases. This makes pragmatic -- given the current maturity of the quantum technologies and given the unavoidable requirement of some sort of local interaction among the qubits to be entangled -- to assume some sort of hierarchy among network nodes, with a specialized super-node -- i.e., the \textit{orchestrator} -- responsible for locally generating and then distributing a multipartite entangled state among the network nodes \cite{EppKamMac-17, AviGusRoz-23, IllCalVis-23,CheIllCac-24}, referred to as \textit{clients}.

By accounting for the discussion about unfeasible dense physical topologies (such as fat tree or leaf-spine) in Sec.~\ref{sec:1}, we consider a sparse topology where the orchestrator is directly connected through physical quantum channels to the network nodes via a star topology, as illustrated in Fig.~\ref{fig:02}.

As defined within the \textit{Research Problem}, we plan to engineer the multipartite entanglement state -- and, more precisely, the graph state -- distributed within the QLAN to overcame the physical topology constraints induced by the quantum hardware underlying QLAN functioning. Clearly, there are two main degrees of freedom underlying the choice of the initial graph state to be generated, distributed and eventually engineered:
\begin{itemize}
	\item[i)] the ``type'' of the graph state, namely the specific structure of the associated graph;
	\item[ii)] the ``dimensions" of the graph state, expressed by the number of qubits: i) retained at the orchestrator and ii) distributed to the clients.
\end{itemize}

\subsubsection{Graph State Type}
Regarding the first degree of freedom, as said, edges of the associated graph are related to some sort of ``entangling interaction'' among the qubits belonging to the composite entangled system. Thus, by over-simplifying, the denser is the graph associated to the graph state, the more challenging is the generation of the corresponding graph state due to the complexity of the underlying multipartite interactions.

In order to confer practicability to our proposal, we consider as elementary state generated at the orchestrator the simplest form of graph states, namely, \textit{linear cluster state}.

\begin{initialstate} 
	Formally, for $n$-qubit \textit{linear cluster state} $\ket{L}$ associated to a linear graph, \eqref{eq:15} reduces to:
	\begin{equation}
		\label{eq:20}
		\ket{L} = \prod_{i = 1}^{n-1} \texttt{CZ}_{(i,i+1)}\ket{+}^{\otimes n}.
	\end{equation}
\end{initialstate}

Indeed, linear cluster state have been already experimentally generated in controlled environments \cite{LeeJeo-23, ShaSha-23}. Furthermore, recently it has been experimentally demonstrated \cite{ThoRusRem-24} that, starting from linear cluster states, it is possible to realize different 2-dimensional graph states, by utilizing properly fusion operations. Thus, our choice of starting from linear cluster states at the orchestrator is not only practical -- being characterized by low complexity -- but it is also not restrictive.

\subsubsection{Graph State Dimensions}
Regarding the second degree of freedom, from its description it is clear that we already made a design choice, i.e., to retain some qubits of the initial state at the orchestrator. In other words, the hierarchy among the nodes is maintained also in the distribution of the multipartite state.

The rationale for this choice is to be able to adapt the resource state -- aka, the graph state generated at the orchestrator -- accordingly to the on-demand traffic requests, without the need of exchanging signaling among the clients. This, in turns, avoids to introduce further delays in a scenario very sensitive to the decoherence.

And, perhaps more importantly, this design choice avoids the need of performing arbitrary quantum operations at the clients, as we prove in the following section. Specifically, the desired artificial topology is built by performing only \text{local} operations on the qubits retained by the orchestrator, starting from the initial distributed state.

 The key role played by the number of qubits retained at the orchestrator is further described in the next subsections, after collecting some definitions.

\subsection{Design Parameters}
\label{sec:3.2}
In this subsection, we map the degrees of freedom related to the entanglement distribution process into a set of design parameters, which allows to completely describe the topology of the graph associated to the graph state.

\begin{definition}[\textbf{Number of orchestration qubits}]
	\label{def:11}
	Given a $n$-qubit graph state $\ket{G}$, the number of orchestration qubits is indicated with:
	\begin{equation}
		\label{eq:21}
		n_o \eqdef |V_o|,
	\end{equation}
	with $V_o = \{o_1, \dots, o_{n_o}\} \subset V$ denoting the subset of vertices of the overall graph $G$ associated with the qubits of the overall graph state $\ket{G}$ retained by the orchestrator.
\end{definition}

\begin{definition}[\textbf{Number of client qubits}]
	\label{def:12}
	Given a $n$-qubit graph state $\ket{G}$, the number of client qubits is indicated with:
	\begin{equation}
		\label{eq:22}
		k \eqdef |V_c|,
	\end{equation}
	with $V_c = \{c_1, \dots, c_{k}\} \subset V$ denoting the subset of vertices of the overall graph $G$ associated with the qubits of the overall graph state $\ket{G}$ distributed to the clients.
\end{definition}

\begin{remark}
	As an example of the key role played by $n_o$ and $k$, let us consider a $k+1$-qubit graph state distributed to $k$ clients. Under these assumptions, we have that the number of qubits at the orchestrator is forced to be $n_o = 1$. Clearly, there exists different \textit{types} of graph states satisfying these constraints, and a simple one -- which also fits with the underlying  physical topology -- is the one associated with a star graph centered at the orchestrator, i.e., $G=(V,E)$ with $E = \big\{ \{o_1, c_i\}_{i=1}^k$ \big\}. Being such a graph state LU-equivalent to a GHZ state \cite{HeiDurEis-06}, it allows  to extract a single EPR pair between any pair of nodes -- i.e., between a couple of clients or between a client and the orchestrator. From this simple example, it appears clear that different choices about $n_o$ and $k$ imply different features of the graph state and, hence, of the associated graph, which in turn determine the clients communication capabilities beyond the physical topology constraint. This will be engineered in the Sec.~\ref{sec:4}.
\end{remark}

\begin{definition}[\textbf{Orchestration qubit: client degree}]
	\label{def:13}
	Given an orchestration vertex $o_i \in V_o$, $k_{o_i,} \leq k$ denotes its ``client'' degree, i.e., the cardinality of its neighborhood $N^c_{o_i} \eqdef N_{o_i} \cap V_c \subset V$, restricted to the vertices associated to qubits distributed to the clients. Formally:
	\begin{align}
		\label{eq:23}
		N^c_{o_i} &\eqdef N_{o_i} \cap V_c = \big\{c_j \in V_c: \{o_i,c_j\} \in E  \big\} \subset V, \\& \nonumber 
  \text{\rm  with,} \,\, k_{o_i,c} = | N^c_{o_i} |. 
	\end{align}
\end{definition}

\begin{figure*}[t!]
	\centering
	\begin{subfigure}{0.45\textwidth}
		\centering
		\includegraphics[width=\textwidth]{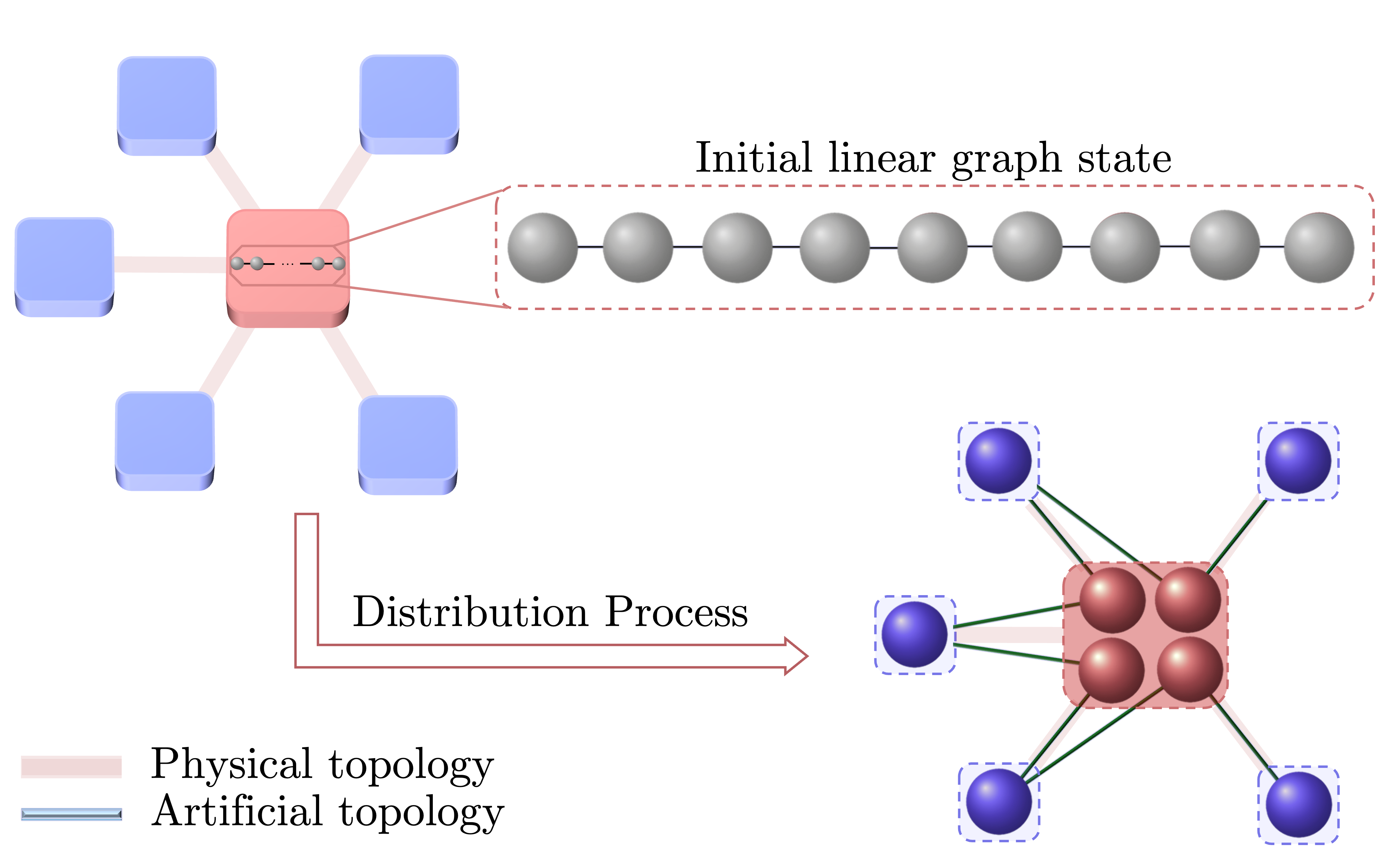}
		\caption{Example of a \textit{chain graph state}, obtained by first generating at the orchestrator a linear cluster state, and then by distributing the entangled qubits to the clients so that any qubit retained at the orchestrator is adjacent to two qubits distributed at two different clients.}
		\label{fig:03-a}
	\end{subfigure}
	\hfill
	\begin{subfigure}{0.50\textwidth}
		\centering
		\includegraphics[width=\textwidth]{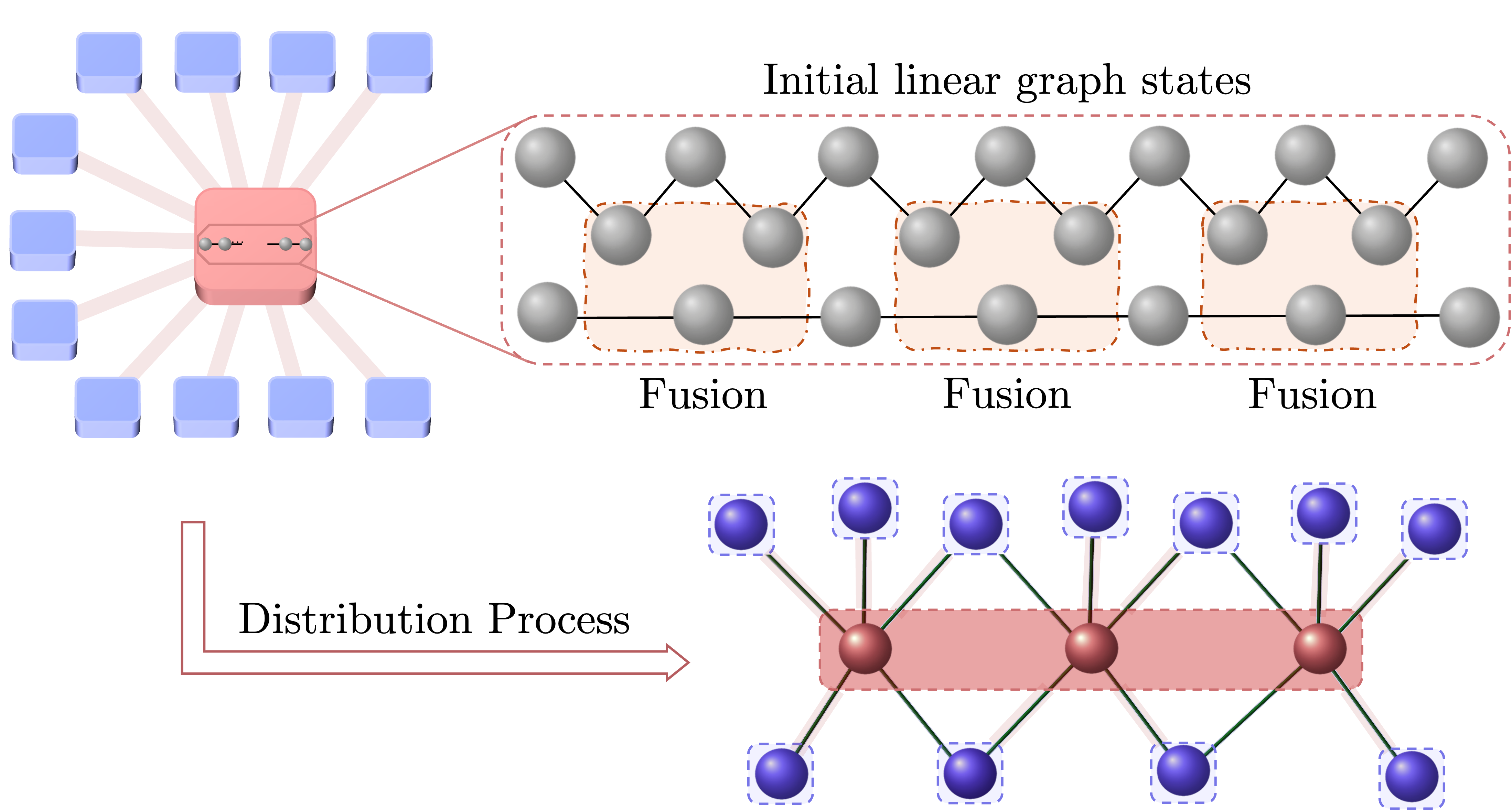}
		\caption{Example of a \textit{generalized tree-like}, obtained by first generating at the orchestrator two linear cluster states, and by wisely performing fusion operations. Then, the qubits are distributed to the $k$ clients, so that any qubit retained at the orchestrator is associated to a vertex, which is adjacent to $k_c$ client vertices.}
		\label{fig:03-b}
	\end{subfigure}
	\caption{Pictorial representation of \textit{chain} and \textit{generalized tree-like graph states}, obtained starting from a linear cluster state.}
	\label{fig:03}
	\hrulefill
\end{figure*}

\begin{definition}[\textbf{Client qubit: $r$-rank bridge}]
	\label{def:14}
	A client vertex $c_i \in V_c$ is defined as ``$r$-rank bridge'' whenever its rank -- i.e., the cardinality of its neighborhood $N^o_{c_i} \eqdef N_{c_i} \cap V_o \subset V$, restricted to vertices associated to qubits retained at the orchestrator -- is $r$, which is greater than one. Formally:
	\begin{align}
		\label{eq:24}
		 N^o_{c_i} &\eqdef N_{c_i} \cap V_o = \big\{o_j \in V_o: \{o_j,c_i\} \in E\big\}\subset V,\\&\nonumber
        \text{\rm with,} \,\, | N^o_{c_i} |=r > 1. 
	\end{align}
\end{definition}

\begin{definition}[\textbf{Orchestration qubit: bridge degree}]
	\label{def:15}
	Given an orchestration vertex $o_i \in V_o$, $0< k^r_{o_i,b} \leq k$ denotes its ``$r$-rank bridge degree'', i.e., the cardinality of its neighborhood $N_{o_i}$, restricted to vertices associated to bridges with rank $r$:
	\begin{equation}
		\label{eq:25}
		k^r_{o_i,b} \eqdef | B^r_{o_i} |, \; \text{\rm with} \; B^r_{o_i} \eqdef \big\{ c_i \in N_{o_i} : | N^o_{c_i} | = r \big\}.
	\end{equation}
\end{definition}

In the following, we denote with the symbols $\bar{k}^r_b$ and $\hat{k}^r_b$ the maximum and minimum values for the $r$-rank bridge degrees among all the orchestration qubits, i.e.:
\begin{align}
	\label{eq:26}
	\bar{k}^r_b & \eqdef \max_{o_i \in V_o} \{ k^r_{o_i,b} \}, \\ 
	\label{eq:27}
	\hat{k}^r_b & \eqdef \min_{o_i \in V_o,} \{ k^r_{o_i,b}\}.
\end{align}

\section{From Physical to Artificial Topology}
\label{sec:4}

Here, we develop the main tools for addressing the research problem introduced in Sec.~\ref{sec:3}:
\begin{quote}
	\textit{bypassing the communication limitations induced by the physical QLAN topology by building artificial topologies -- such as \textit{bus} and \textit{(enhanced) ring} topologies -- interconnecting nodes at run-time, accordingly to the traffic demand.}
\end{quote}

We highlight that motivations and interest for these artificial topologies are not arbitrary. Indeed, the graph states associated to these topologies exhibit an application value, representing the main resources for measurement-based quantum information processing and computation \cite{BriRau-01,RauBri-01,BarBirBom-23}. Furthermore, limiting the analysis to these artificial topologies is not restrictive, since it is possible to build different topologies starting from the considered ones, as experimentally proved \cite{ThoRusRem-24}.

\subsection{Distributed State Design}
\label{sec:4.1}

As pointed out in Sec.~\ref{sec:3.1}, we consider -- as elementary multipartite states generated at the orchestrator -- linear cluster states since they are experimentally-feasible and they can be merged to obtain more complex states.

Yet, there exists different design choices in choosing: i) the specific final graph state (obtained by combining the elementary states) to be distributed and ii) the specific distribution pattern of the individual qubits of the final state to the nodes of the QLAN. We summarize our design choices in the following.

\begin{designprinciples}
	The generation and distribution process of the $n$-qubit graph state $\ket{G}$ through a QLAN with $k$ clients is performed so that the graph $G=(V,E)$ associated to $\ket{G}$ satisfies the following conditions:
	\begin{align}
		\label{eq:28}
		& \text{i)} \;\;\;	V = V_o \cup V_c \; \wedge \; V_o \cap V_c = \emptyset \\
		\label{eq:29}
		& \text{ii)} \;\;	|V_o| > 1 \; \wedge \; |V_c | = k \\ 
		\label{eq:30}
		& \text{iii)} \;	\forall \{a,b\} \in E : a \in V_o \; \wedge \; b \in V_c \\
		\label{eq:31}
		& \text{iv)} \; \;	k_{o_i,c} = k_c \, \forall \, o_i \in V_o, \; \wedge \exists := r':\\ 
		& \quad \quad \; \begin{cases}
				 k^{r'}_{o_i,b} \neq  0, \, \forall \, o_i \in V_o \\
				k^{r'}_{b,o_i} = \bar{k}^{r'}_b = \bar{k}_b \vee k^{r'}_{b,o_i} = \hat{k}^{r'}_b = \hat{k}_b, \, \forall o_i \in V_o\\
			\end{cases} \nonumber
	\end{align}
\end{designprinciples}

The first constraint is quite axiomatic, forcing the orchestrator to distribute the qubits of the multipartite entangled state, either to clients or to itself, as a local resource for forcing the artificial connectivity among the clients toward the topology of interest, as we prove in the following.

As for the second constraint, it allows us to consider the worst-case scenario from a communication perspective: each client receives only one qubit of the multipartite entangled state for fulfilling the on-demand traffic requests. 

The third constraint forces the graph state to exhibit ``vertical'' edges among the two different hierarchy levels represented by $V_o$ and $V_c$ -- i.e., between orchestrator and clients vertices -- rather than `horizontal'' edges within the same level. This design choice is key to ``remotely'' tune the artificial connectivity among the clients by only manipulating the qubits at the orchestrator, as proved in the following. Furthermore, we note that the absence of horizontal edges \textit{enforces a two-colorable structure} \cite{HeiDurEis-06,HeiEisBri-04} on the distributed graph state. In fact, a graph $G=(V,E)$ is two-colorable if the set of vertices $V$ can be partitioned into two subsets  so that there exist no edge in $E$ between two vertices belonging to the same subset. For our modeling, the final graph state $G=(V,E)$ is also denoted as $G=(V_o, V_c, E)$ whenever we want to empathize the two-colorable property. 

Finally, as for the fourth constraint, it enforce a recursive and regular structure within the graph underlying the graph state after the distribution, in the light of practicability. Accordingly, we require\footnote{We note that, having a fixed values for $k_{o_i,c}$ is not restrictive, since this condition can be easily satisfied by introducing fictitious clients during the entanglement generation process.} that the client degree is $k_{o_i,c} = k_c$ for all the orchestration qubits. Furthermore, we requires that all the bridges have the same rank -- say $r'$ adjacent orchestrator vertices -- and that each orchestrator is adjacent to either $\bar{k}_b$ or $\hat{k}_b$ bridges\footnote{Hence, it results $k_c \geq \bar{k}_b \geq \hat{k}_b$ by definition.}. Accordingly, the final graph state exhibit a recursive and regular structure, as shown in Fig~\ref{fig:03}, consisting in a recursive topology built by concatenating elementary constituents represented by the star subgraph $G[\dot{N}_{o_i}]$ induced by the closed neighborhoods $\dot{N}_{o_i}$ of the orchestrator vertices. In this topology, the maximum number of bridges $\bar{k}_b$ is exhibited by intermediate orchestration qubits, whereas the the minimum number of bridges is $\bar{k}_b$ is exhibited by the two orchestration qubits at the edges of the structure.

Stemming from the above four design principles, we design two different type of graph states to be distributed within the QLAN -- both characterized by $2$-rank bridges -- referred to as \textit{chain graph state} and \textit{generalized tree-like graph state}, and formally defined in the following.

\begin{chaingraph}
	A $n$-qubit `chain'' graph state can be distributed through a QLAN with $k$ clients by retaining $n_o = k-1$ qubits at the orchestrator and by setting $k_c = 2$, $\bar{k}_b=2$ and $\hat{k_b} = 1$. Accordingly, the associated graph $G=(V,E)$ satisfies the following:
	\begin{align}
		\label{eq:32}
		 V &=  V_o \cup V_c, \; \text{\rm with } V_o \eqdef \{ o_i \}_{i=1}^{k-1} \; \wedge \; V_c \eqdef \{ c_i \}_{i=1}^{k}, \\
		\label{eq:33}
		E &= \bigcup_{i=1}^{k-1} \big\{ \{o_i,c_i\}, \{o_i,c_{i+1}\} \big\}.
	\end{align}
\end{chaingraph}

\begin{remark}
	A chain graph state can be straightforwardly obtained from a linear graph state $\ket{L}$ as depicted in Fig.~\ref{fig:03-a}. In a nutshell, it is sufficient to generate a $(2k-1)$-qubit linear cluster state at the orchestrator, and to wisely distribute $k$ qubits to the clients, so that any qubit retained at the orchestrator is associated to a vertex, which is adjacent to two client vertices corresponding to the qubits distributed to two different clients. 
\end{remark}

\begin{treelikegraph}
	A $n$-qubit ``generalized tree-like'' graph state can be distributed through a QLAN with $k$ clients by retaining $n_o = \frac{k-\hat{k_b}}{k_c-\hat{k_b}}$ qubits at the orchestrator for arbitrary values of $k_c,\bar{k}_b,\hat{k_b} \in \mathbb{N^{+}}$. Accordingly, the associated graph $G=(V,E)$ satisfies the following:
	\begin{align}
		\label{eq:34}
		 V &=  V_o \cup V_c, \; \text{\rm with } V_o \eqdef \{ o_i \}_{i=1}^{n_o} \; \wedge \; V_c \eqdef \{ c_i \}_{i=1}^{k} \\
		\label{eq:35}
		E &= \bigcup_{i=1}^{n_o} \big ( \{o_i\}\times N_{o_i} \big).
	\end{align}
	%where $N_{o_i} = \{ c_{\tilde{i}+1}, c_{\tilde{i}+2}, \ldots, c_{\tilde{i}+k_c} \}$ with $\tilde{i} = (i-1) (k_c - \hat{k_b})$.
\end{treelikegraph}

From \eqref{eq:35}, it is evident that the graph associated to the generalized tree-like graph, after the distribution process, appears as the concatenation  of $n_o$ star subgraphs, defined in Def.~\ref{def:05}, each having $k_c$ edges -- $G[\dot{N}_{o_i}]= (\dot{N}_{o_i},\{o_i\}\times N_{o_i})$ with $i \in \{1, \ldots, n_o\}$ -- induced by $\dot{N}_{o_i}$ and with $\{o_i\}_{i =1}^{n_o}$ as star vertexes of the subgraphs. The concatenation utilizes as anchor vertices  the $\bar{k_b}$ bridges for internal orchestrator qubits and as anchor vertices the $\hat{k_b}$ bridges for the $2$ external orchestrator qubits.
Clearly, similar considerations can be made for the chain graph, by looking at \eqref{eq:33}.

\begin{remark}
	As depicted in Fig.~\ref{fig:03-b} for $k_c=5$, $\bar{k_b}=4$ and $\hat{k_b}=2$, a generalized tree-like graph state can be obtained by generating at the orchestrator two linear cluster states and by wisely performing fusion operations at the orchestrator. After that, the qubits are distributed to the $k$ clients, so that any qubit retained at the orchestrator is associated to a vertex, which is adjacent to $k_c$ client vertices corresponding to the qubits distributed to the $k_c$ different clients. It is evident from Figs.~\ref{fig:06} and ~\ref{fig:07} that the structure of the graph resembles a tree. This consideration induced us to label it as “generalized tree-like” graph state
\end{remark}

Stemming from the concept of bridge, we can now provide the last definition, utilized in the theoretical analysis.

\begin{definition}[\textbf{Client proximity}]
	\label{def:16}
	Given two clients $c_i, c_j \in V_c \subset V$ with $i \neq j$, their proximity $d(c_i,c_j)$ is defined as the number of bridges belonging to the shortest path, connecting the two clients within the graph $G = (V, E)$, plus one. Formally::
	\begin{equation}
		\label{eq:36}
		d(c_i,c_j) = 1+\big| \big\{ a \in \, \overline{p}_{\{c_i, c_j\}} : a \in V_c \, \wedge |N^o_a| > 1 \big\} \big|,
	\end{equation}
	with $\overline{p}_{\{c_i, c_j\}}$ denoting the shortest path among all the possible paths defined in Def.~\ref{def:09}.
\end{definition}

We highlight that two clients $c_i, c_j \in V_c \subset V$  adjacent to the same orchestration qubit in the designed resource states have the minimum possible value of proximity, i.e., $ d(c_i,c_j) = 1$.

\subsection{From Physical Star Topology to Artificial Bus Topology}
\label{sec:4.2}

Here, we prove in Lemma~\ref{lem:01} how to engineer a \textit{chain graph state} distributed through the QLAN so that all the clients are eventually interconnected by an an artificial bus topology, i.e., a linear graph among the vertices associated to qubits stored at the clients.

\begin{lemma}
	\label{lem:01}
	By distributing a $(2 k -1)$-qubit chain graph state through the QLAN, an artificial bus topology interconnecting $k$ clients can be obtained by performing $n_c=(k-1)$ local $\sigma_y$-Pauli measurements of the qubits retained at the orchestrator.
	\begin{IEEEproof}
		Please refer to Appendix~\ref{a:sec1} in the Supplementary Material. 
	\end{IEEEproof}
\end{lemma}

\begin{figure}[t!]
	\centering
		\input{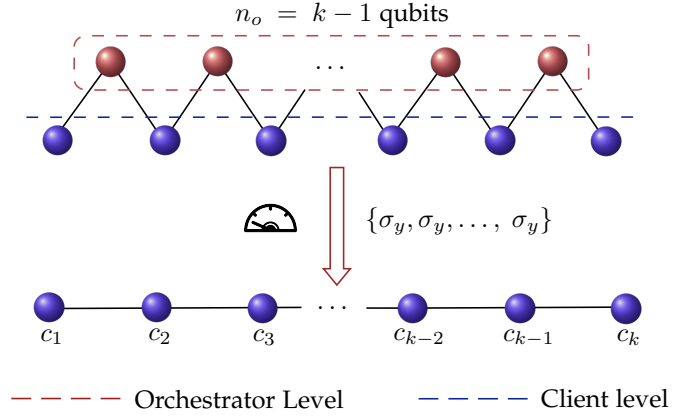}
		\caption{Generation of an artificial bus topology among the $k$ clients of the QLAN starting starting from a $(2k-1)$-qubit chain state. The artificial topology is obtained by (wisely) measuring each qubit retained at the orchestrator, according to Lem.~\ref{lem:01}.}
		\label{fig:04}
		\hrulefill
\end{figure}

As depicted in Fig.~\ref{fig:04}, the results of Lemma~\ref{lem:01} imply that is possible to build an artificial topology directly interconnecting clients with artificial links -- even if they are not physically connected in the physical topology -- by exploiting only local operations at the orchestrator. From a communication engineering perspective, this is valuable since the orchestrator can tune the artificial connectivity for dynamically satisfying the client traffic patterns \textit{after} -- rather than \textit{before} -- the entanglement distribution process has been completed. 

\begin{remark}
	Clearly, the orchestrator could have distributed a bus (linear) graph state since the beginning as initial state. Yet, whenever the client communication needs at run-time would involve pairs of clients that happen to be distant within the linear topology, the distributed state should be further processed to be adapted to those needs. This requires a sequence of quantum operations and classical coordination/signaling at and between the clients -- as an example, entanglement swapping at the intermediate node(s) \cite{} -- for satisfying such needs, inducing so further overhead and delays in a scenario very sensitive to the decoherence. On the contrary, our framework allows to build at run-time the most suitable topology, without the need of additional signaling nor quantum operations at the clients, by executing only local operations on the qubits retained by the orchestrator, starting from the initial multipartite state.
\end{remark}

To provide a deeper insight of the ability of our framework to dynamically adapt to traffic demands, let us consider the extraction of EPR pairs between couple of clients as the final communication task, as instance for performing qubit transmission via quantum teleportation. To this aim, the artificial bus topology enables the simultaneous extraction of $\lfloor{\frac{k}{2}}\rfloor$ EPR pairs \cite{HeiDurEis-06}, depending on the identities of the clients aiming at communicate each other. One could be induced to believe that this task would require some sort of cooperation from the clients. This is true in general, but our framework allows to achieve the same result by exclusively acting locally at the orchestrator, as proved with the following lemma.

\begin{lemma} 
	\label{lem:02}
	By distributing a $(2 k -1)$-qubit chain graph state through the QLAN among $k$ clients, then up to $\lfloor{\frac{k}{2}}\rfloor$ EPR pairs can be obtained by performing $n_c$ Pauli-measurements on the qubits retained at the orchestrator.
	\begin{IEEEproof}
		Please refer to Appendix~\ref{a:sec2} in the Supplementary Material. 
	\end{IEEEproof}
\end{lemma}

Stemming from this result, it is worth to discuss a parallel with respect to classical LAN topologies. Specifically, this capability to fulfill in parallel up to $\lfloor{\frac{k}{2}}\rfloor$ different qubit transmissions has no counterpart in classical LANs resembling the same topology, such as the classical \textbf{bus topology}. Indeed, in such a classical case, a single medium -- e.g., a coaxial cable -- is shared among all the LAN nodes, which allows only one communication per use of the channel. Furthermore, although cost-effective and easy to deploy, the classical bus topology introduces a point-of-failure vulnerability: whenever the bus fails, the entire LAN experiences service disruption \cite{MetBog-76}. On the contrary, the persistency of the graph state described by the artificial bus topology is indeed $\lfloor{\frac{k}{2}}\rfloor$ \cite{HeiDurEis-06,RiePol-11}. The persistency -- namely, the minimum number of qubits that need to be measured to guarantee that the resulting state is separable \cite{IllCalMan-22} -- indicates the robustness of a multipartite state against losses or accidental measurements of a qubit, which destroy entanglement. In this light, the persistency can be seen as a quantum equivalent of resistance to point-of-failure vulnerability, mentioned above. Thus, also by accounting for this communication metric, the artificial bus topology represents an improvement with respect to the classical world.

\begin{figure}[t!]
	\centering
		\input{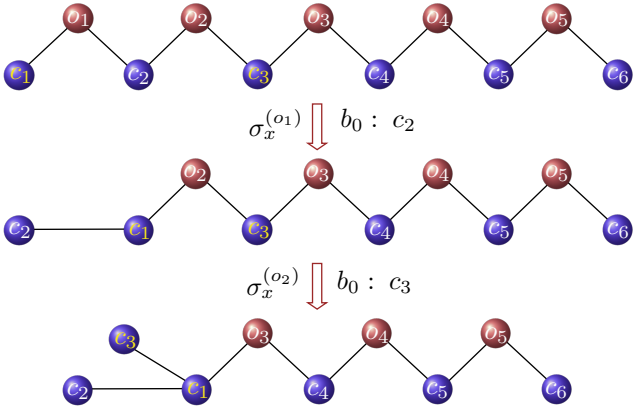}
		\caption{Entanglement rolling: generation of an artificial link between two clients $c_i$ and $_cj$ starting from a $(2k-1)$-qubit chain state. In the example, the clients to be interconnected within the artificial topology are $c_1$ and $c_3$,- whose proximity distant in the initially distributed chain state is $d(c_1,c_3) = 2$.}
		\label{fig:05}
		\hrulefill
\end{figure}

The above mentioned capability of the artificial topology to dynamically adapt to traffic demands by processing qubits retained at the orchestrator is further stressed by the following result, where $d(c_i,c_j)$ denotes the proximity between clients $c_i$ and $c_j$ defined in Def.~\ref{def:16} .

\begin{lemma} 
	\label{lem:03} \textbf{(Entanglement Rolling)}
		By distributing a $2 k -1$-qubit chain graph state through the QLAN among $k$ clients, an artificial link inter-connecting two clients $c_i,c_j \in V_c, i \neq j$ can be built by performing $d(c_i,c_j)$ $\sigma_x$ Pauli-measurements on the qubits retained at the orchestrator and associated to vertices belonging the the shortest path $\overline{p}_{c_i,c_j}$ connecting $c_i$ and $c_j$.  
	\begin{proof}
		\renewcommand{\qedsymbol}{}
		Please refer to Appendix~\ref{a:sec3} in the Supplementary Material. 
	\end{proof}
\end{lemma}

A pictorial representation of the results of Lemma~\ref{lem:03} is reported in Fig.~\ref{fig:05}. There, the two clients $c_1$ and $c_3$ -- which are neither physical connected nor virtually connected in the initial distributed multipartite state -- are eventually connected by an artificial link. For this, it suffices to perform $2=d(c_1,c_3)$ $\sigma_x$-Pauli measurements on specific orchestration qubits. We named the effects induced by Lemma~\ref{lem:03} on the topology as \textit{entanglement rolling} to highlight the roller effects on the client artificial connections. It is worthwhile to emphasize that the result of Lemma~\ref{lem:03} is not equivalent to extract EPR pairs from the overall multipartite state. It rather goes in the direction of properly manipulating and adapting the artificial topology, by relaying on the orchestration qubits to effectively adapt to the traffic-demands and at the same time to save in terms of communication overhead.  

\begin{remark}
 As detailed in the appendix, the key role played by the bridges in the designed resource states is to act as anchors in the artificial topology. Specifically, bridges act as anchors capable of connecting different sub-nets within the overall topology due to their connections with multiple orchestration qubits. And indeed, by exploiting the bridges, communication opportunities are facilitated  among clients adjacent to different orchestration qubits, by properly manipulating them as proved in Lemmas~\ref{lem:01} and \ref{lem:03}
\end{remark}

\subsection{From Physical Star Topology to Artificial Enhanced Ring Topology}
\label{sec:4.3}

\begin{figure}[t!]
	\centering
	\input{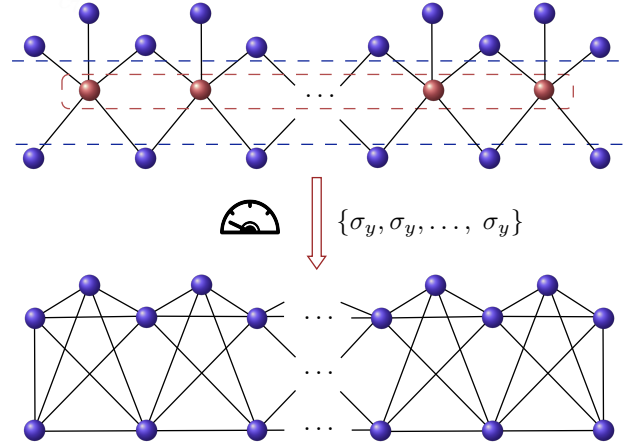}
	\caption{Generation of an enhanced ring topology among the $k$ clients of the QLAN starting from a $n$-qubit tree-like state with $k_c=5$, $\hat{k_b}=2$ and $\bar{k}_b=2\,\hat{k_b}=4$. The artificial topology is obtained by (wisely) measuring each qubit retained at the orchestrator, according to Lem.~\ref{lem:04}.}
	\label{fig:06}
	\hrulefill
\end{figure}

The on-demand capability of the artificial quantum topology to adapt to the different traffic demands is further enhanced by engineering a graph state with a denser connectivity -- i.e., the generalized tree-like graph -- by paying a price consisting in the higher complexity of the state generation process.

Here, we first prove in Lemma~\ref{lem:04} how to obtain an artificial enhanced ring topology among the clients, starting from the \textit{generalized tree-like graph state}. This artificial topology is referred to as ``enhanced ring topology'', due to its structure, which resembles the classical ring topology augmented with additional edges among the clients. 

\begin{figure}[t!]
    \centering
        \input{Tikz/Fig7.tex}
        \caption{Generation of an enhanced ring topology among the $k$ clients of the QLAN starting from a $(\frac{3}{2} k - 1)$-qubit tree-like graph state, with $k_c = \bar{k}_b = 4$, $\hat{k_b}=2$ and $\overline{k}_b=2 \hat{k_b}=4$. In particular, the artificial topology is obtained by (wisely) measuring each qubit retained at the orchestrator, according to Lem.~\ref{lem:04}.}
        \label{fig:07}
    \hrulefill
\end{figure}
    
\begin{lemma}
    \label{lem:04}
	By distributing a $n$-qubit generalized tree-like graph state through the QLAN, then an artificial enhanced ring topology interconnecting the $k$ clients, characterized by an edge set with cardinality equal to $n_o {k_c \choose 2} - 2(n_o - 1){\hat{k}_b \choose 2}$, can be obtained by performing $n_o= \frac{k-\hat{k}_b}{k_c-\hat{k}_b}$ local Pauli $\sigma_y$-measurements on the orchestrator qubits.
    \begin{IEEEproof}
        Please refer to Appendix~\ref{a:sec3} in the Supplementary Material. 
    \end{IEEEproof}
\end{lemma}

\begin{figure*}[t!]
    \centering
    \begin{subfigure}[b]{\textwidth}
        \centering
        \input{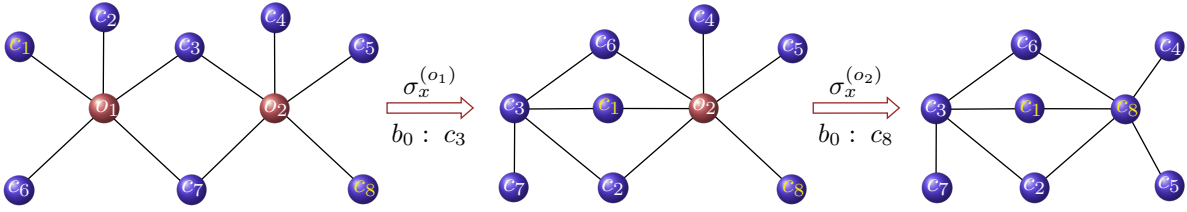}
 \caption{Example of entanglement rolling with generalized tree-like topology and $k_c = 5$, $n_o = 2$ and $\hat k_b = 2$. In this example, clients $c_1$ and $c_8$ have a proximity $d(c_1,c_8) = 2$.}
        \label{fig:08-a}
     \hrulefill
     \end{subfigure}
    \vfill
    \begin{subfigure}[b]{\textwidth}
        \centering
        \input{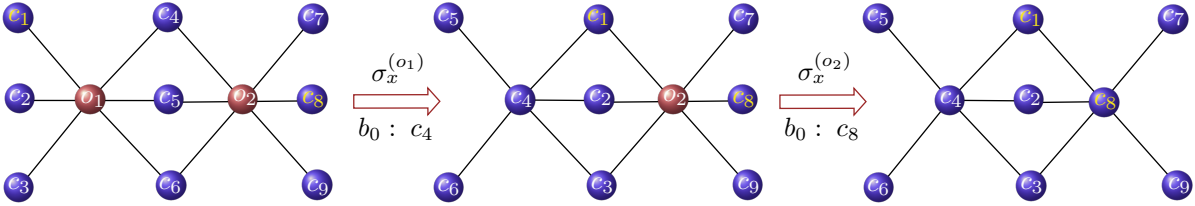}
        \caption{Example of entanglement rolling with generalized tree-like topology and $k_c = 6$, $n_o = 2$ and $\hat k_b = 3$. In this example, clients $c_1$ and $c_8$ have a proximity $d(c_1,c_8) = 2$.}
        \label{fig:08-b}
      \hrulefill
  \end{subfigure}
    \vfill
    \begin{subfigure}[b]{\textwidth}
        \centering
        \input{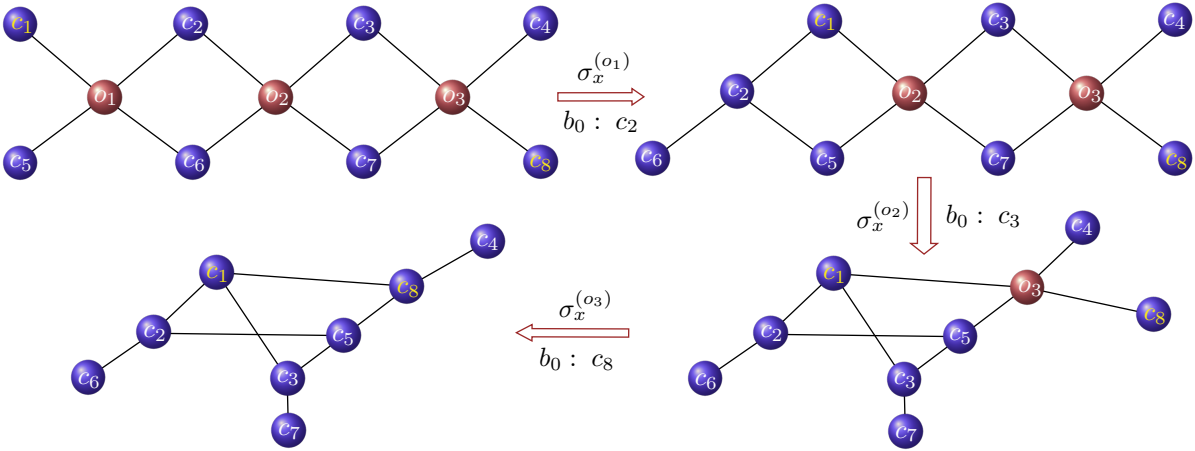}
        \caption{Example of entanglement rolling with generalized tree-like topology and $k_c = 4$, $n_o = 3$ and $\hat k_b = 2$. In this example, clients $c_1$ and $c_8$ have a proximity $d(c_1,c_8) = 3$.}
        \label{fig:08-c}
    \end{subfigure}
    \caption{Pictorial representation of the entanglement rolling effects on generalized-tree like states, accordingly to Lemma~\ref{lem:05}.}
    \label{fig:08}
    \hrulefill
\end{figure*}

The result of Lemma~\ref{lem:04} implies that the final neighborhood of a client -- i.e., the number of artificial links generated at each client by measuring the qubits at the orchestrator -- depends on $k_c$, which in turns is lower bounded by the number of bridges, i.e., $k_c \geq \bar{k}_b$. This is evident by comparing Fig.~\ref{fig:06} and Fig.~\ref{fig:07}, characterized by two different values of $k_c$. Indeed, it is valuable to observe that for a generalized tree-like graph with $k_c$ edges for each orchestration vertex equal to the minimum one -- i.e., $k_c=\bar{k_b}=2\hat{k_b}$ as represented in Fig.~\ref{fig:07} -- the cardinality of the edge set of the artificial enhanced ring topology simplifies to $\hat{k}_b((n_o+1) \hat{k_b} -1)$.

We emphasize that we named the built artificial topology as ``enhanced ring topology'', since its structure resembles the shape of a classical ring topology augmented with additional links among the nodes. The concept of artificial enhanced ring topology represents a remarkable quantum counterpart of the classical ring topology. Indeed, in classical ring topologies, each node communicates with exactly two neighboring nodes. Data travels along the ring, passing from one device to the next one until reaching its destination \cite{BuxClo-83,StrNor-87}. Despite offering significant advantages over classical bus (such as simpler routing algorithms) topologies, a classical ring topology cannot tolerate the failure of neither the bus nor any single node, and it poses significant deployment challenges when it comes to network expansion.

Conversely, artificial enhanced ring topologies do not
constraint each node to communicate with exactly two neighboring nodes, since each client has multiple artificial pathways to connect with the selected client destination. Thus, the network flexibility and adaptability to the traffic demands is even further increased with respect to the artificial bus topology. 
Furthermore, as we will prove in the next section, the persistency of the graph state associated to the 
artificial enhanced ring topology is $n_o$, which is greater than $1$. Thus, as for the artificial bus topology, the enhanced ring topology overcomes the single point-of-failure inherent in classical ring scenarios. In other words, if one or more qubits are lost or measured, the remaining clients in the network can still communicate by utilizing the remaining artificial links within the topology. Furthermore, the ability to dynamically reconfigure paths based on the redundant artificial connections of the artificial topology enhances the overall reliability and adaptability of the quantum network. 

It is also interesting to observe that, in the case of artificial topologies enabled by entanglement, the network expansion can be easily achieved by increasing the size of the multipartite state distributed within the QLAN, hence by overcoming another challenge of the classical world.  

The anchor role played by the bridges and highlighted for the chain graph resource state can be further stressed for the generalized tree-like resource, as proved in the following. 

\begin{lemma} \textbf{(Entanglement Rolling)}
    \label{lem:05}
    By distributing a $n$-qubit generalized tree-like graph state though the QLAN among $k$ clients, an artificial link interconnecting two clients $c_i,c_j \in V_c, i \neq j$ can be built by performing $d(c_i,c_j)$ $\sigma_x$-Pauli measurements on the qubits retained at the orchestrator and associated to vertices belonging the shortest path $\overline{p}_{c_i,c_j}$ connecting $c_i$ and $c_j$.
    \begin{IEEEproof}
        Please refer to Appendix~\ref{a:sec5} in the Supplementary Material. 
    \end{IEEEproof}
\end{lemma}

A pictorial representation of the results of Lemma~\ref{lem:05} is
reported in Fig.~\ref{fig:08}. In such a figure, for different versions of the generalized tree-like topology, non-adjacent clients $c_1$ and $c_8$ are finally connected by an artificial link,
i.e, they become neighbor in the artificial topology, by
performing $d(c_1, c_8)$ $\sigma_x$-Pauli measurements on the orchestration qubits. Also for this case, we named the effects induced
by Lemma~\ref{lem:05} on the topology, as entanglement rolling to
highlight the roller effects on the client artificial connections.

\subsection{Enhanced Ring: Quantifying Entanglement}
\label{sec:4.4}
One of the aspect worthwhile of further analysis is to quantify the entanglement in enhanced ring topologies.

To this aim, a widely used approach consists in evaluating the Schmidt measure  $E_s(\ket{G})$ \cite{HeiDurEis-06}. However, even if the Schmidt measure stands as an important tool for quantifying the entanglement of a quantum state, it can be very hard to calculate, since it requires the decomposition of the quantum state in the LU-equivalent quantum state characterized by the smallest number of superposed terms. 

Indeed, since the enhanced ring is obtained through $\sigma_y$ measurements on the orchestration qubits of a generalized-tree like state, the two-colorable structure of the original state is not assured \cite{HeiDurEis-06}. Stemming from this observation, in order to quantify the entanglement within an artificial enhanced ring topology, the preliminary result in Lemma~\ref{lem:06} is needed.

\begin{lemma}
    \label{lem:06}
    An artificial enhanced ring topology shared among $k$ clients -- obtained by engineering a $n$-qubit generalized tree-like graph characterized by $k_c>\bar{k}_b$ according to Lemma~\ref{lem:04} -- is Local-Clifford (LC) equivalent to a $k$-qubit generalized tree-like graph state with the same values for $n_c$ and the same number of bridges of the original tree-like graph, but with a number of edges per orchestration qubit given by $k'_c=k_c-1$ and with a number of clients equal to $k'=k-n_c$.
    \begin{IEEEproof}
        Please refer to Appendix.~\ref{a:sec6} in the Supplementary Material.
    \end{IEEEproof}
\end{lemma}

Although the condition $k_c =\bar{k}_b$ is not captured by Lemma~\ref{lem:06}, this is not restrictive since we can always assure $k_c > \overline{k}_b$, by adding fictitious nodes as discussed in Sec.~\ref{sec:4.1}.

\begin{lemma}
    \label{lem:07}
    The  Schmidt measure of the graph state associated to an artificial enhanced ring topology shared among $k$ clients -- obtained by engineering a $n$-qubit generalized tree-like graph characterized by $k_c>\bar{k}_b$ according to Lemma~\ref{lem:04} --admits a closed form expression as follows:
    \begin{equation}
        \label{eq:37}
        E_S(\ket{G_{er}}) = n_o.
    \end{equation}
    \begin{IEEEproof}
        Graph states that are LC equivalent are characterized by the same Schmidt measure. Hence,  by accounting for the result of Lemma~\ref{lem:06}, it is  sufficient to determine the Schmidt measure of the LC-equivalent generalized tree-like graph state with $k_c > \bar k_b$. To this aim, we observe that this graph state is a two-colorable graph, for which lower and upper bounds are known \cite{HeiDurEis-06}. By observing that the size of the minimum vertex cover is $n_o$, being $n_o< k'$, as well the rank of the submatrix $\Gamma_{AB}$ of the adjacency matrix of the overall graph state, the proof follows, since the upper and lower bounds coincide. 
    \end{IEEEproof}
\end{lemma}

According to Lemma~\ref{lem:07}, we can extract $n_o$ EPR pairs from an enhanced ring. But, differently from plain ring graph states with the same number of qubits, the freedom in selecting the identities of the pairs is higher. Indeed, the $\frac{n_o}{2}k_c(k_c-3)-n_ok\hat{k}_b \frac{\hat{k}_b -2}{k-\hat{k}_b }$ additional edges with respect to a $k$-qubit ring increases the degrees of freedom in selecting the pairs of nodes that eventually will share an EPR pairs. Thus, this type of topology is suitable for communication scenarios characterized by highly-variable traffic patterns. 

Finally, form the above lemma, it results also that the peristency of the artificial enhanced ting topology is $n_o$, which, as mentioned above, overcome the single-point failure of classical ring topologies. 

\section{Conclusions} 
\label{sec:5}
In this paper, we have introduced and modeled the pivotal role played by multipartite entanglement within Quantum Local Area Network (QLAN) topology. Specifically, we have shown that the engineering of the artificial network topology enabled by multipartite entanglement can be performed on-demand, according to the communication needs, by exploiting only local Pauli measurements at the node responsible for multipartite entanglement generation and distribution. To this aim, we proved that it is possible, by starting from a physical star topology and by wisely manipulating multipartite entanglement, to build different artificial topologies. We hope that this work, by proposing a new perspective on the concept of quantum LANs, will fuel the interest of the community towards QLANs as building block for the future Quantum Internet.

\bibliography{bibliography}
\bibliographystyle{ieeetr}

\newpage

\begin{appendices}

\section{Proof of Lemma 1} 
\label{a:sec1}
\begin{IEEEproof}
Accordingly to Sec.~\ref{sec:2}, the action of a $\sigma_y$-Pauli measurement at the orchestrator on the qubit associated to vertex $o_i$ is equivalent to the local complementation of the graph at vertex $o_i$, followed by the deletion of $o_i$ from the graph. Thus accordingly to \eqref{eq:18}, it results:
\begin{align}
    \label{eq:38}
    &\tau_{o_i}(G) - o_i = \\
    &= (V \setminus \{o_i\} \; , \;  (E \cup N_{o_i}^{2}) \setminus E_{N_{o_i}} \setminus E_{o_i}), \nonumber
\end{align}
where $E_{o_i}\eqdef \{o_i\} \times {N_{o_i}}$.

From the definition of a $n$-qubit chain graph state, it is easy to recognize that $E_{N_{o_i}}$ is an empty set, $E_{o_i}\eqdef \{o_i\} \times {N_{o_i}}=\big\{ \{o_i,c_i\}, \{o_i,c_{i+1}\}\big\}$ and that $N_{o_i}^2 = \{c_i,c_{i+1}\}$. Thus by performing a $\sigma_y$-Pauli measurement on the $i$-th vertex at the orchestrator, \eqref{eq:38} is equivalent to:
\begin{align}
    \label{eq:39}
    &\tau_{o_i}(G) - o_i = \\
    &= (V \setminus \{o_i\} \; , \;  E \cup \{c_i,c_{i+1}\} \setminus \big\{\{o_i,c_i\},\{o_i,c_{i+1}\} \big\}). \nonumber
\end{align}
By reasoning as above and by accounting for \eqref{eq:38} and \eqref{eq:39}, $n_o$ $\sigma_y$-Pauli measurements at the $n_o$ orchestrator vertices lead to the graph $ \tilde{G}_y ^{(n_o)} = (V^{(n_o)},E^{(n_o)})$ with:
\begin{align}
    \label{eq:40}
    V^{(n_o)} & = V \setminus \bigcup_{i=1}^{k-1} \{o_i\} = V \setminus V_o,\\
    \label{eq:41}
    E^{(n_o)} & = \big\{ \{c_i,c_{i+1}\}_{i=1}^{k-1} \big\}.
\end{align}
The proof follows by recognizing that  the resulting graph $\tilde{G}_y ^{(n_o)}$ exhibits a $k$-qubit BUS topology among all the initial $k$ clients.
\end{IEEEproof}

\section{Proof of Lemma 2} \label{a:sec2}
\begin{IEEEproof}
Let us consider the projection operations on $i$-th orchestration qubit defined as follows:
 \begin{align}
       \label{eq:42}
       P^{(o_i)} = 
        \begin{cases}
            P_{y}^{(o_i)} \otimes \mathbb{I}^{\otimes(n-1)} \qquad &\text{if $i$ is odd}, \\
            P_{z}^{(o_i)} \otimes \mathbb{I}^{\otimes(n-1)} \qquad &\text{if $i$ is even},
        \end{cases}
\end{align}
where $P$ is the projection operator associated with the $\sigma_z$ or $\sigma_y$-Pauli measurements -- depending on the index -- applied on qubit $o_i$ and $\mathbb{I}$ is the identity operator applied on the rest of the qubits. 

Accordingly to Sec.~\ref{sec:2}, the action of a $\sigma_z$-Pauli
measurement on the qubit associated to
vertex $o_i$ is equivalent to the deletion of vertex $o_i$ from the
graph. Whereas, the action of a $\sigma_y$-Pauli measurement on the qubit associated to vertex $o_i$ is equivalent to the local complementation of the graph at vertex $o_i$, followed by the deletion of $o_i$ from the graph.

From this, it is evident that the resulting graph obtained via $\sigma_z$-Pauli measurement on qubit $o_i$ and the resulting graph obtained via $\sigma_y$-Pauli measurement on qubit $o_i$ are characterized by the same vertex set, while they differ in the edge sets. More into detail, the projection operator on the $i$-th orchestration qubit leads to the following graph: 
    \begin{align}
        \label{eq:43}
        \tilde G ^{(i)} = \begin{cases}
            \tilde G_y ^{(i)} = (V_y ^{(i)}, E_y ^{(i)}) \qquad \text{if $i$ is odd}, \\ 
            \tilde G_z ^{(i)} = (V_z ^{(i)}, E_z ^{(i)}) \qquad \text{if $i$ is even},
        \end{cases}
    \end{align}
with: 
\begin{equation}
    \label{eq:44}
    V_y ^{(i)} = V_z ^{(i)}=V\setminus\{o_i\},
\end{equation}

\begin{equation}
    \label{eq:45}
    E_y ^{(i)} = (E \cup \{c_i,c_{i+1}\} )\setminus \big\{\{o_i,c_i\},\{o_i,c_{i+1}\} \big\},
\end{equation}

\begin{equation}
    \label{eq:46}
    E_z ^{(i)} = E  \setminus \big\{\{o_i, c_i\}, \{o_i,c_{i+1}\} \big\}.
\end{equation}

Thus by performing $n_o$ measurements according to \eqref{eq:42}, the vertex set of the resulting graph is given by the unmeasured vertices, i.e. the clients, while, by accounting for \eqref{eq:44} and \eqref{eq:45}, the set of edges is given by the links between two consecutive clients whose smaller index is odd. Formally:
    \begin{equation}
        \label{eq:47}
        \tilde G^{(n_o)} = \bigg(
        \underbrace{V \setminus \bigcup_{i=1}^{n_o}\{o_i\}}_{V^{(n_o)}} \; , \; 
        \underbrace{\bigcup_{i=0}^{\lceil \tfrac{n_o}{2}\rceil -1}\{c_{2i+1},c_{2i+2}\}}_{E ^{(n_o)}} \bigg).
    \end{equation}

Thus, the graph state associated to the graph $\tilde G^{(n_o)}$ can be written as:
    \begin{equation}
        \label{eq:48}
        \ket{\tilde G} = \bigotimes_{i=0}^{\lceil \tfrac{n_o}{2}\rceil} \ket{K_2},
    \end{equation}
where $\ket{K_2}$ is the two-qubit fully connected graph state, in \eqref{eq:08}, which is LU equivalent to a Bell state. Remarkably, for the chain graph state topology, we have that $n_o = k-1$, therefore $\lceil \tfrac{n_o}{2}\rceil = \lceil \tfrac{k-1}{2}\rceil = \lfloor \tfrac{k}{2} \rfloor$. This completes the proof.
\end{IEEEproof}

\section{Proof Lemma 3}
\label{a:sec3}
As indicated in Sec.~\ref{sec:2}, the $\sigma_x$-measurement on a qubit corresponding to orchestrator vertex $o_i$ is equivalent to perform the following sequence of graph operations
\begin{equation}        
\label{eq:49}
     \tau_{b_0} \left( \tau_{o_i}\big(\tau_{b_0}(G) )-o_i  \right),
\end{equation}
with $b_0$ an arbitrary neighbor of $o_i$. 
By accounting for the structure of the chain graph and by Def.~\ref{def:09}, it results that the shortest path  $\overline{p}_{\{c_i,c_j\}}$ connecting $c_i$ and $c_j$, is composed by $d(c_i,c_j)$-orchestrator vertices and $d(c_i,c_j)-1$ bridges, being $d(c_i,c_j)$ their proximity. 
By accounting for this consideration, the proof follows by setting the neighbors $\{b_0\}$ involved in the first $d(c_i,c_j)-1$ $\sigma_x$-measurements on the orchestrator vertices equal to the identities of the $d(c_i,c_j)-1$ bridges belonging to $\overline{p}_{\{c_i,c_j\}}$, and the last $b_0$ -- of the $\sigma_x$-measurement on the last orchestrator qubit -- is set equal to the client $c_j$.

By proceeding step-by-step, the first $\sigma_x$-measurement is performed on $o_i$, with $b_0=c_{i+1}$. If $d(c_i,c_j) > 1$, $c_{i+1}\neq c_j$ is a bridge, otherwise $c_{i+1}=c_j$ and the proof directly follows. Thus, \eqref{eq:49} can be re-written as:
\begin{equation}
    \label{eq:50}
    \tau_{c_{i+1}} \left( \tau_{o_i}\big(\tau_{c_{i+1}}(G) )-o_i  \right),
\end{equation}
with 
\begin{align}
    \label{eq:51}
    \tau_{c_{i+1}}(G) = \big( V, \underbrace{(E \cup N_{c_{i+1}}^2) \setminus E_{N_{c_{i+1}}}}_{E^{'}} \big),
\end{align}
and $N_{c_{i+1}}^2 = \{o_i,o_{i+1}\}$.
Moreover, we have that:
\begin{equation}
    \label{eq:52}
    \tau_{o_i}(\tau_{c_{i+1}}(G))= \big( V, \underbrace{(E^{'} \cup N_{o_i}^2) \setminus E_{N_{o_i}}}_{E^{''}} \big),
\end{equation}
where the set $N_{o_i}^2$ is as follows: 
\begin{equation}
    \label{eq:53}
    N_{o_i}^2 = 
    \begin{cases}
    \big\{ \{c_i, c_{i+1}\}, \{ c_i, o_{i+1}\}, \{c_{i+1}, o_{i+1}\} \big\} \quad &\text{if } i < n_o,\\
     \{ c_i, c_{i+1}\} \quad &\text{if } i = n_o, \\
    \end{cases}
\end{equation}
and the set $E_{N_{o_i}}$ is given by:
\begin{equation}
    \label{eq:54}
    E_{N_{o_i}} = \begin{cases}
        \{c_{i+1}, o_{i+1}\} \quad &\text{if } i < n_o, \\
        \emptyset \quad &\text{if } i = n_o.
    \end{cases}
\end{equation}
Accordingly, the edge set $E^{''}$ includes the edges $\{\{c_i, o_{i+1}\},\{c_i,c_{i+1}\},\{c_i,o_i\}, \{c_{i+1},o_i\}\}$. Stemming from this, it results that the deletion of vertex $o_i$ in \eqref{eq:50} leads to the resulting graph: 
\begin{equation}
    \label{eq:55}
    \tau_{o_i}(\tau_{c_{i+1}}(G)) - {o_i}= \big( V \setminus \{o_i\}, \underbrace{E^{''} \setminus E_{o_i}}_{E^{'''}} \big),
\end{equation}
with $E_{o_i}=\big\{ \{o_i, c_i\},\{ o_i, c_{i+1}\},\{o_i,o_{i+1}\}\big\} $. Thus, $E^{'''}$ includes the edges $\{\{c_i, o_{i+1}\},\{c_i,c_{i+1}\}\}$. We note that the actions of the former graph operations lead to the scenario in which $c_i$ is the only neighbor of $c_{i+1}$. Therefore, the last graph operation in \eqref{eq:50}, i.e. $\tau_{c_{i+1}}(\tau_{o_i}(\tau_{c_{i+1}}(G)) - {o_i})$,  does not change the graph. 

The above results show that by performing a $\sigma_x$-measurement on $o_i$ and by choosing as support node $b_0$ the bridge $c_{i+1}$, i.e. $b_0=c_{i+1}$, the overall effect is to create a direct edge between $c_i$ and $c_{i+1}$ and between $c_i$ and $o_{i+1}$, while $c_{i+1}$ looses its bridge role, meaning that it looses the edge with $o_{i+1}$. Thus $c_i$ and $c_{i+1}$ are swapped in their artificial topology positions, reducing so the proximity between $c_i$ and $c_j$. From this description is already evident the ``rolling'' effect, mentioned in Sec.~\ref{sec:4}. 

By reasoning as above and by performing the other $(d(c_i,c_j)-2)$ $\sigma_x$-measurements on $d(c_i,c_j)-2$ orchestrator qubits $\{o_k\}_{k=i+1}^{i+d(c_i,c_j)-2}$, $c_i$ is progressively swapped in its topological position with the $d(c_i,c_j)-2$ bridges, belonging to the shortest path $\overline{p}_{\{c_i,c_j\}}$. Hence, at the last measurement stage on $o_{i+d(c_i,c_j)-1}$, $c_i$ exhibits an edge with $o_{i+d(c_i,c_j)-1}$ which in turns has an edge with $c_j$. Thus by choosing as support node $b_0=c_j$ and by reasoning as above, the proof follows. 

\section{Proof of Lemma 4} 
\label{a:sec4}
 Without loss of generality, in the following, we restrict our attention on artificial topologies characterized by $k_c > \bar k_b$. The proof can be carried similarly also for the easier case in which $k_c = \bar{k}_b$.
 
 Specifically, the proof follows by adopting a similar reasoning as in Lem.~\ref{lem:01}: first local complementations of the graph $G$ -- associated to the generalized tree-like graph state -- at vertices $\{o_i\}_{i=1}^{n_o}$ are performed and then, each of the aforementioned complementation is followed by the deletion of $\{o_i\}$ from the resulting graph, as indicated in \eqref{eq:18}.

Accordingly to Sec~\ref{sec:4}, the graph $G = (V,E)$ associated to a $n$-qubit generalized tree-like graph state can be expressed through star subgraphs. Formally: 
\begin{equation}
    \label{eq:56}
    G= \bigcup_{i=1} ^{n_o} G[\dot N_{o_i}] = \bigcup_{i=1} ^{n_o} (\dot N_{o_i}, \underbrace{\{o_i\} \times N_{o_i}}_{E_{o_i}}),
\end{equation}
with $N_{o_i}$ the neighborhood associated to an arbitrary orchestration vertex $o_i$.
To carry the proof, it is useful to explicit the neighborhood $N_{o_i}$.

To this aim, we introduce a labeling for the clients based on the splitting of the clients into two groups with increasing numbering from left to right. Specifically, client vertices are assumed to be placed in two separate groups, named \textit{up} and \textit{down}, as follows.
\begin{align}
    \label{eq:57}
    &V_c = V_{up} \cup V_{down}, \, \wedge \; V_{up}  \cap V_{down} = \emptyset, \\
    \label{eq:58}
   & V_{up} = \big\{  \{c_i\}_{i=1}^{k_{f}} \big\}, \\
    \label{eq:59}
    &V_{down} = \big\{ \{c_j\}_{j={k_{f}}+1}^{k} \big\},
\end{align} 
with $k_{f}$ denoting an offset value defined as: $k_{f} \eqdef \lceil \tfrac{k_c}{2} \rceil n_o - \lceil \tfrac{\hat k_b}{2} \rceil(n_o-1)$. 
We introduce also other two offset parameters characterizing a certain orchestration qubit $o_i$, as follows:
\begin{align}
    \label{eq:60}
    & k_{f,o_i}^{\rm up}=
        (\lceil \tfrac{k_c}{2} \rceil - \lceil \tfrac{\hat k_b}{2} \rceil - 1)(i-1)  \\
    \label{eq:61}
       &k_{f,o_i}^{\rm down}= (\lfloor \tfrac{k_c}{2} \rfloor - \lfloor \tfrac{\hat k_b}{2} \rfloor - 1)(i-1). 
\end{align}

By accounting for \eqref{eq:58}, \eqref{eq:59} and \eqref{eq:60}, \eqref{eq:61}, the neighborhood $N_{o_i}$ in \eqref{eq:56} of each $o_i$ can expressed:
\begin{align} 
    \label{eq:62}
    N_{o_i} &= \bigg\{ \{c_j\} \in V_{up}: j=i+k_{f,o_i}^{\rm up},\dots,i+k_{f,o_i}^{\rm up}+\\
    &+(\lceil \tfrac{k_c}{2} \rceil -1)\bigg\} \cup \bigg\{ \{c_l\} \in V_{down}: l=i+k_{f} + \nonumber \\
    &+k_{f,o_i}^{\rm down},\dots,i+k_{f}+k_{f,o_i}^{\rm down} +(\lfloor\tfrac{k_c}{2} \rfloor -1)\bigg\} \nonumber.
\end{align}

Stemming from the above, it results that the action of a $\sigma_y$-Pauli
measurement at the orchestrator
vertex $o_1$ is equivalent to the local complementation of the
graph at vertex $o_1$, followed by the deletion of $o_1$ from the
graph:
\begin{align}
    \label{eq:63}
    &\tilde G^{(1)} = \tau_{o_1}(G) - o_1 = \\
    &= (\underbrace{V \setminus \{o_1\}}_{\eqdef V^ {(1)}} \; , \;  \underbrace{[ (E \cup N_{o_1}^{2}) \setminus E_{N_{o_1}}] \setminus E_{o_1}}_{\eqdef E^{(1)}}), \nonumber
\end{align}
with $E_{N_{o_1}} = \emptyset$, as a consequence of the definition of generalized tree-like state. Accordingly, a $\sigma_y$-Pauli measurement at the orchestrator
vertex $o_1$ leads to a new graph where all the clients originally in $N_{o_1}$ in \eqref{eq:63} are fully interconnected, including the clients with bridge role.

This consideration allows us to highlight that at the next measurement step, when a $\sigma_y$-Pauli
measurement is performed at the orchestrator vertex $o_2$, $E_{N_{o_2}} $ is not anymore an empty set. To provide the expression of $E_{N_{o_i}}$ at the arbitrary measurement step at at the orchestrator vertex $o_i$, it is useful to introduce the edge set of the bridges connected to a given orchestrator  vertex $o_i$:
\begin{equation}
    \label{eq:64}
    E_{B_{o_i}} = \big\{ \{c_i,c_j\}: c_i, c_j \in B_{o_i}, i \neq j \big\} \subset E^{(i)},
\end{equation}
where $B_{o_i}$ is defined in \eqref{eq:25}. 
Accordingly, in the $i$-th measurement step, the action of a $\sigma_y$-Pauli measurement at the orchestrator vertex $o_i$ leads to the graph:
\begin{align}
        \label{eq:65}
        \tilde G^{(i)} = \tau_{o_i}(\tilde G^{(i-1)}) - o_i  
        = (V^{(i)},E^{(i)} \big)
    \end{align}
with vertex and edge sets -- $V^{(i)}$, $E^{(i)}$ -- depending on the vertex and edge sets -- $V^{(i-1)}$, $E^{(i-1)}$ -- in the previous measurement step:
    \begin{align}
        \label{eq:66}
        V^{(i)} &= V^{(i-1)} \setminus \{o_i\} \\
        \label{eq:67}
        E^{(i)} &= (E ^{(i-1)} \cup N_{o_i}^{2}) \setminus E_{N_{o_i}}\setminus E_{o_i}.
    \end{align}
The set $E_{N_{o_i}}$ in \eqref{eq:67} contains the links created between bridges adjacent to the orchestration vertices $o_i$ and $o_{i-1}$:
\begin{align}
    \label{eq:68}
    &E_{N_{o_i}} = 
    \begin{cases}
    \emptyset   \quad &\text{ if } i=1, \\
    E_{B_{o_i}} \cap  E_{B_{o_{i-1}}}  \quad &\text{ otherwise},
    \end{cases}
\end{align}
and $N_{o_i}^{2}$ is given by:
\begin{equation}
    \label{eq:69}
    N_{o_i}^{2} = \big\{ \{c_j, c_l\}: c_j,c_l \in N_{o_i}, j \neq l \big\}.
\end{equation}
Accordingly, at each $\sigma_y$ measurement step, a new complete subgraph $K[N_{o_i}] = (N_{o_i},N_{o_i}^{2})$ induced by the neighbors of $o_i$ is created. Such a complete subgraph is reduced by the deletion of the edges between bridge vertices belonging to $E_{Bo_i} \cap  E_{Bo_{(i-1)}}$, i.e., bridges adjacent to $o_i$ and $o_{i-1}$. 
It is convenient to note that the complete subgraph $K[N_{o_i}]$ can be equivalently obtained by the union of star subgraph associated to each client in $N_{o_i}$:
    \begin{equation}
        \label{eq:70}
        K[N_{o_i}] = \bigcup_{\forall c_i \in N_{o_i}} G[\dot N_{c_i}],
    \end{equation}
where $G[\dot N_{c_i}]$ is defined in \eqref{eq:10} in Def.~\ref{def:05} as:  
    \begin{equation}
        \label{eq:71}
        G[\dot {N_{c_i}}] = (\underbrace{N_{o_i}}_{\dot N_{c_i}} \; , \underbrace{\{c_i\} \times (N_{o_i} \setminus \{c_i\}) }_{\{c_i\} \times N_{c_i}}) = (\dot V_{c_i} \; , \; \dot E_{c_i}).
    \end{equation} 
Thus, the overall artificial topology, obtained after $n_o$ $\sigma_y$-Pauli measurements, is given by the union of all the star subgraphs associated to each client, without the edges between internal bridges. Formally:
    \begin{equation}
        \label{eq:72}
        \tilde{G}_y^{(n_o)}  = \bigg( 
        \underbrace{
        V \setminus \bigcup_{i=1}^{n_o}\{o_i\}}_{V^{(n_o)}
        } \; , \;  
        \underbrace{
        \bigcup_{i=1}^{k}\dot E_{{c_i}} \setminus  \bigcup_{i=1}^{n_o}
        (E_{Bo_i} \cap  E_{Bo_{(i-1)}})
        }_{E ^{(n_o)}} \bigg). 
    \end{equation} 

From the above, it results that for each external orchestration qubit, namely $o_i=o_1$ and $o_i=o_{n_o}$, $k_c \choose 2$ new edges are created as consequence of the complementation, while at the next measurement step, exactly $\hat k_b \choose 2$ links are deleted among the bridges belonging to $E_{Bo_i} \cap  E_{Bo_{(i-1)}}$. For all the other (internal) orchestration qubits, the number of deleted edges is doubled. Formally, the cardinality of the edge set can be written as:
\begin{align}
    \label{eq:73}
    \nonumber
    |E^{(n_o)}|& = 2 \bigg[{k_c \choose 2} - {\hat k_b \choose 2}\bigg] + (n_o - 2)\bigg[{ k_c \choose 2} - 2{\hat k_b \choose 2} \bigg] = \\&=
    n_o {k_c \choose 2} - 2(n_o - 1){\hat k_b \choose 2}.
\end{align}
This complete the proof.

\section{Proof Lemma 5}
\label{a:sec5}
Similarly to Lemma~\ref{lem:03}, the proof follows by setting the neighbors $\{b_0\}$ involved in the first $d(c_i,c_j)-1$ $\sigma_x$-measurements on the orchestrator vertices equal to the identities of the $d(c_i,c_j)-1$ bridges belonging to $\overline{p}_{\{c_i,c_j\}}$, and the last $b_0$ -- of the $\sigma_x$-measurement on the last orchestrator qubit -- is set equal to the client $c_j$.

For the sake of notation simplicity we assume that $c_i$ is a client associated to orchestrator $o_i$. Otherwise, a re-labeling of the client $c_i$ is assumed.  This is not restrictive due to the symmetry of the structure of the generalized tree-like graph described in Sec.~\ref{sec:4}. Accordingly, if $d(c_i,c_j)>1$, the $b_0$ of the first $\sigma_x$-measurement has to be set equal to one of the bridges in $B_{o_i}$, as for instance, $b_0=c_{i+k_{f,o_i}^{\rm up}+(\lceil \tfrac{k_c}{2} \rceil -1)}$, with $k_{f,o_i}^{\rm up}$ defined in \eqref{eq:60}.  In the following for the sake of clarity, we denote with $\ell_{o_i}\eqdef k_{f,o_i}^{\rm up}+(\lceil \tfrac{k_c}{2} \rceil -1)$. As observed in the proof of Lemma~\ref{lem:03}, the effect of the $\sigma_x$-measurement is to swap the positions within the artificial topology between $c_i$ and the support node $c_{i+ \ell_{o_i}}$, in terms of edges with the orchestrator vertices. 

By proceeding step-by-step, the first $\sigma_x$-measurement, performed on $o_i$ with $b_0=c_{i+ \ell_{o_i}}$, has the effect of modifying the neighbor of $c_{i+ \ell_{o_i}}$ as follows:
\begin{equation}
\label{eq:74}
    N_{c_{i+ \ell_{o_i}}} = ( N_{o_i} \setminus \{c_{i + \ell_{o_i}}\} ).
\end{equation}
Accordingly to \eqref{eq:74}, $c_{i+ \ell_{o_i}}$ is not anymore a bridge for $o_{i+1}$.
It is also interesting to note that the aforementioned behavior is common to each bridge of the orchestrator vertices $o_i$ and $o_{i+1}$ in the initial graph. In other words, each vertex belonging to $(B_{o_i} \cap B_{o_{i+1}})$ with $i < n_o$, has the vertex $c_{i + \ell_{o_i}}$ as the only neighbor, after the measurement, loosing so its bridge role. 
On the contrary, client $c_i$ assumes the role of bridge for $o_{i+1}$:
\begin{equation}
    \label{eq:75}
    N_{c_i} = 
    \begin{cases}
        \big\{ c_{i + \ell_{o_i}}, o_{i+1} \big\} \quad &\text{if } i < n_o,\\
        \{c_{i + \ell_{o_i}}\} \quad &\text{if } i = n_o.
    \end{cases}
\end{equation}
The aforementioned  behavior is also exhibited by the other clients originally -- before the $\sigma_x$-measurement -- in $N_{o_i}$ and not in $B_{o_i} \cap B_{o_{i+1}}$.

By accounting for the above, the overall effect of the $\sigma_x$-measurement on $o_i$ is to create an artificial link between $c_i$ and $o_{i+1}$, by highlighting again the ``rolling'' effect mentioned in Appendix~\ref{a:sec3}. The proof follows, by reasoning as above. Specifically, by performing the other $(d(c_i,c_j)-2)$ $\sigma_x$-measurements on $d(c_i,c_j)-2$ orchestrator qubits, $c_i$ is progressively swapped in its topological position with the $d(c_i,c_j)-2$ bridges, belonging to the shortest path $\overline{p}_{\{c_i,c_j\}}$. Hence, at the last measurement stage on $o_{i+d(c_i,c_j)-1}$, $c_i$ exhibits an edge with $o_{i+d(c_i,c_j)-1}$ which in turns has an edge with $c_j$. Thus by choosing as support node $b_0=c_j$ and by reasoning as above, the proof follows.

\section{Proof Lemma 6}
\label{a:sec6}
Accordingly to Def.~\ref{def:10}, two $k$-qubit graph states are LC equivalent iff e corresponding graphs are related by a sequence
of local complementations. 

Let us consider a $k$-quibit generalized tree-like graph state whose associated graph $G=(V,E)$ is:
\begin{align}
       \label{eq:76}
        V &=  V_o \cup V_c \eqdef\big\{ \{ o_i \}_{i=1}^{n_o} \cup \{ c_i \}_{i=1}^{k'} \big\} \\
        \label{eq:77}
        E &= \bigcup_{i=1}^{n_o} E_{o_i},
    \end{align}  
where $E_{o_i}\eqdef \{o_i\}\times N_{o_i}$ is the edge set, with cardinality $k'_o=k_c-1$, associated to  $G[\dot{N}_{o_i}]$ and $k'=k-n_o$. 
By locally complementing $G$ at vertex $o_1$, one has:
\begin{equation}
    \label{eq:78}
    \tau_{o_1}(G) = \bigg( V, (E \cup N_{o_1}^2) \setminus E_{N_{o_1}} \bigg),
\end{equation}
where $E_{N_{o_1}}$ is the empty set. In the remaining $n_o-1$ local complementations at vertices $\{o_i\}_{i=2}^{n_o}$, by reasoning as in rom Appendix~\ref{a:sec4}, it results that $E_{N_{o_i}}$ is given by:
\begin{equation}
    \label{eq:79}
    E_{N_{o_i}} =
        E_{B_{o_i}} \cap  E_{B_{o_{(i-1)}}}.
\end{equation}
Accordingly, at the last local complementation on vertex $o_{n_o}$, the associate graph can be written as:
\begin{equation}
    \label{eq:80}
    G^{(n_o)} = 
    \bigg(
    V,
    \bigcup_{i=1}^{n_o} (E^{(i-1)} \cup N_{o_i}^2) \setminus \bigcup_{i=2}^{n_o} (E_{B_{o_i}} \cap  E_{B_{o_{(i-1)}}})
    \bigg),
\end{equation}
with $E^{(0)}=E$, being the originally edge set before the first complementation at node $o_1$.
From \eqref{eq:80}, it is easy to recognize that:
\begin{equation}
    \label{eq:81}
    (E^{(i-1)} \cup N_{o_i}^2) = \bigcup_{i=1}^{k'} \dot E_{c_i} \bigcup_{i=1}^{n_o} E_{o_i}= \bigcup_{i=1}^{k'+n_o} \dot E'_{c_i}.
\end{equation}
By substituting \eqref{eq:81} in \eqref{eq:80}, one has:
\begin{equation}
    \label{eq:82}
    G^{(n_o)} = 
    \bigg(
    V,
    \bigcup_{i=1}^{k'+n_o} \dot E'_{c_i}\setminus \bigcup_{i=2}^{n_o} (E_{B_{o_i}} \cap  E_{B_{o_{(i-1)}}})
    \bigg).
\end{equation}
The proof follows, by recognizing that  \eqref{eq:82} is equivalent to \eqref{eq:72} in Appendix~\ref{a:sec4}. In other words, an artificial enhanced ring with $k=k'+n_o$ clients is obtained. 

\end{appendices}

\end{document}